\documentclass[10pt,conference]{IEEEtran}
\IEEEoverridecommandlockouts
\usepackage{cite}
\usepackage{amsmath,amssymb,amsfonts}
\usepackage{graphicx}
\usepackage{textcomp}
\usepackage{xcolor}
\usepackage[hyphens]{url}
\usepackage{fancyhdr}
\usepackage{hyperref}
\usepackage{enumitem}
\usepackage[]{hyperref}
\usepackage{adjustbox}
\usepackage{graphicx}
\usepackage{subcaption}
\usepackage{caption}
\usepackage{float} 
\usepackage{longtable}
\usepackage{multirow}
\usepackage{amsmath} 
\usepackage{tikz}
\usepackage{titlesec}
\usepackage{listings}
\usepackage{algorithm}
\usepackage[noend]{algpseudocode}  
\usepackage{stfloats}
\usepackage{soulutf8}  
\soulregister\cite7    
\soulregister\sysname7
\soulregister\autoref7
\sethlcolor{yellow} 
\usepackage{xcolor}  
\usepackage{booktabs}
\usepackage{marginnote}

\newcommand{\xg}[1]{{\color{blue} #1}}
\renewcommand{\xg}[1]{#1}

\newcommand{\hpcayear}{2026}
\newcommand{\sysname}{\emph{LiveUpdate}}

\newcommand{\hpcasubmissionnumber}{1058}
\title{Near-Zero-Overhead Freshness for Recommendation Systems via Inference-Side Model Updates}

\def\hpcacameraready{} 

\newcommand\hpcaauthors{First Author$\dagger$ and Second Author$\ddagger$}
\newcommand\hpcaaffiliation{First Affiliation$\dagger$, Second Affiliation$\ddagger$}
\newcommand\hpcaemail{Email(s)}

\renewcommand{\hpcaauthors}{
    Wenjun Yu\IEEEauthorrefmark{2},
    Sitian Chen\IEEEauthorrefmark{2},
    Cheng Chen\IEEEauthorrefmark{3},
    Amelie Chi Zhou\IEEEauthorrefmark{2}\textsuperscript{*}
}

\renewcommand{\hpcaaffiliation}{
    \IEEEauthorrefmark{2}Hong Kong Baptist University \quad
    \IEEEauthorrefmark{3}ByteDance
}

\renewcommand{\hpcaemail}{
    \{cswjyu, csstchen, amelieczhou\}@comp.hkbu.edu.hk,
    chencheng.kit@bytedance.com
}

\author{
  \ifdefined\hpcacameraready
    \IEEEauthorblockN{\hpcaauthors{}}
      \IEEEauthorblockA{
        \hpcaaffiliation{} \\
        \hpcaemail{}
      }
  \else
    \IEEEauthorblockN{\normalsize{HPCA \hpcayear{} Submission
      \textbf{\#\hpcasubmissionnumber{}}} \\
      \IEEEauthorblockA{
        Confidential Draft \\
        Do NOT Distribute!!
      }
    }
  \fi 
}

\fancypagestyle{camerareadyfirstpage}{%
  \fancyhead{}
  
  \fancyhead[C]{
    \ifdefined\aeopen
    \parbox[][12mm][t]{13.5cm}{\hpcayear{} IEEE International Symposium on High-Performance Computer Architecture (HPCA)}    
    \else
      \ifdefined\aereviewed
      \parbox[][12mm][t]{13.5cm}{\hpcayear{} IEEE International Symposium on High-Performance Computer Architecture (HPCA)}
      \else
      \ifdefined\aereproduced
      \parbox[][12mm][t]{13.5cm}{\hpcayear{} IEEE International Symposium on High-Performance Computer Architecture (HPCA)}
      \else
      \parbox[][0mm][t]{13.5cm}{\hpcayear{} IEEE International Symposium on High-Performance Computer Architecture (HPCA)}
    \fi 
    \fi 
    \fi 
    \ifdefined\aeopen 
      \includegraphics[width=12mm,height=12mm]{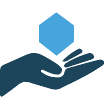}
    \fi 
    \ifdefined\aereviewed
      \includegraphics[width=12mm,height=12mm]{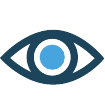}
    \fi 
    \ifdefined\aereproduced
      \includegraphics[width=12mm,height=12mm]{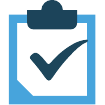}
    \fi
  }
  \fancyfoot[C]{}
}
\begin{document}
\maketitle

\ifdefined\hpcacameraready 
  \thispagestyle{camerareadyfirstpage}
  \pagestyle{empty}
\else
  \thispagestyle{plain}
  \pagestyle{plain}
\fi

\newcommand{\hpcaheight}{0mm}
\ifdefined\eaopen
\renewcommand{\hpcaheight}{12mm}
\fi

\maketitle

\begingroup
\renewcommand\thefootnote{} 
\footnotetext{
  \rule{2cm}{0.4pt}\vspace{0.4ex}
  
  \textsuperscript{*}Corresponding author.
}
\endgroup


\begin{abstract}
\label{re:abstract}
Deep Learning Recommendation Models (DLRMs) underpin personalized services but face a critical freshness-accuracy tradeoff due to massive parameter synchronization overheads. Production DLRMs deploy decoupled training/inference clusters, where synchronizing petabyte-scale embedding tables (EMTs) causes multi-minute staleness, degrading recommendation quality and revenue. We observe that (1) inference nodes exhibit sustained CPU underutilization (peak $\leq$20\%), and (2) EMT gradients possess intrinsic low-rank structure, enabling compact update representation. We present \sysname{}, a system that eliminates inter-cluster synchronization by co-locating Low-Rank Adaptation (LoRA) trainers within inference nodes. \sysname{} addresses two core challenges: (1) \textit{dynamic rank adaptation} via singular value monitoring to constrain memory overhead ($<$2\% of EMTs), and (2) \textit{NUMA-aware resource scheduling} with hardware-enforced QoS to eliminate update-inference contention (P99 latency impact $<$20ms). Evaluations show \sysname{} reduces update costs by 2$\times$ versus delta-update baselines while achieving higher accuracy within 1-hour windows. By transforming idle inference resources into freshness engines, \sysname{} delivers online model updates while outperforming state-
of-the-art delta-update methods by 0.04–0.24\% in accuracy. 
\end{abstract}

\vspace*{-3ex}
\section{INTRODUCTION}
\label{sec:introduction}

Deep Learning Recommendation Models (DLRMs) are fundamental infrastructure powering personalized user experiences across major online platforms, from social media feeds to e-commerce recommendations. Their operational scale is staggering: within Meta's infrastructure alone, DLRMs consume over 50\% of training cycles and 60\% of inference resources, representing a dominant share of modern datacenter computation~\cite{gupta2020architectural,acun2021understanding}. This resource footprint stems from their complex architecture, which combines dense neural networks (e.g., MLPs, Transformers) with massive embedding tables (EMTs) storing high-dimensional representations of users and items. Industrial deployments now reach petabyte scale (e.g., Meta's 96TB Model-F and Kuaishou's 200TB+ Persia~\cite{mudigere2022software,lian2022persia,jiang2019xdl,zhao2023recd}), creating unprecedented systems challenges.

Production DLRMs deploy a decoupled architecture, where \emph{training clusters} continuously update parameters using streaming user interactions, and \emph{inference clusters} serve predictions using the latest parameters synced via centralized parameter servers (see Figure~\ref{fig:dlrm_system}).
This separation optimizes specialized hardware (e.g., GPUs for parameter update in training cluster and CPUs for EMT storage in inference cluster), but introduces severe synchronization overhead across clusters. For example, syncing just 10\% of a 200TB EMT (20TB) over 100GbE networks takes over 26 minutes.
During this delay, inference nodes operate with \emph{stale parameters}, directly degrading recommendation quality.

Maintaining model freshness is crucial for revenue-critical services~\cite{zhang2023survey,he2023dynamically,wang2024rap}, 
as the model recommendation accuracy decays rapidly without updates (e.g., \autoref{fig:freq_acc}). 
Industry studies confirm that even a 0.1\% accuracy drop can translate to millions in lost revenue~\cite{eisenman2022check,lin2024understanding}, while update delays of more than 5 minutes can measurably reduce user engagement~\cite{ktena2019addressing}. Therefore, production systems require near-real-time parameter updates to ensure that models remain fresh and accurate. However, synchronizing multi-terabyte EMTs across clusters over commodity networks incurs significant latencies, far exceeding acceptable freshness windows. This highlight the need for innovative solutions that can simultaneously balance freshness, accuracy, and overhead.

Existing solutions struggle to resolve this tension. \emph{Delta-based} updates~\cite{xie2020kraken,sima2022ekko,liu2022monolith,wei2022gpu} synchronize only the changed parameters since the last update, thus reducing the data volume compared to a full sync. However, as quantified in Figure~\ref{fig:criteo_tb}, the update volume remains massive (e.g., exceeding 10\% of EMTs even in short 10-minute windows). This still translates to multi-minute synchronization delays and unacceptable staleness. 
\emph{Prioritization-based} updates (e.g., QuickUpdate~\cite{matam2024quickupdate}) go a step further by transferring only a small subset of ``{important}'' changed parameters (e.g., top 5-10\% by gradient magnitude).
However, this strategy still incurs multi-minute delays, and the magnitude-based heuristic omits semantically critical but low-gradient updates, directly leading to accuracy degradation.
Ultimately, these prior approaches are fundamentally limited by the training-inference decoupled architecture, which intrinsically creates an inter-cluster bandwidth bottleneck. 

By examining production traces, we identify two overlooked opportunities that motivate a rethinking of DLRM update mechanisms.
\emph{First, inference nodes exhibit sustained and severe CPU underutilization.} Our analysis reveals that CPUs in inference nodes are constantly underutilized during serving, with peak utilization reaching only 20\% (\autoref{fig:cpu_utilization}). This creates significant headroom for local computation at inference clusters. 
\emph{Second, embedding gradients possess a strong intrinsic low-rank structure.}
Our analysis using Principle Component Analysis (PCA) reveals that over 80\% of parameter update variance can be captured by a tiny fraction (e.g., $<$5\%) of principal components (\autoref{fig:pca_tables}). This allows updates to be represented compactly as low-rank matrices, with approximation error bounded by the Eckart-Young theorem~\cite{eckart1936approximation}.
Together, these insights enable a paradigm shift: we can co-locate lightweight training within inference nodes, leveraging idle CPUs to compute compact, low-rank updates. This eliminates inter-cluster synchronization entirely while preserving model accuracy.

Exploiting these opportunities, this paper presents \sysname{}, a system that embeds Low-Rank Adaptation (LoRA) trainer~\cite{hu2022lora} directly within inference nodes.
Each node performs continuous, local updates to its embeddings using recently cached interaction data. Instead of transmitting raw parameter deltas, it computes and applies updates via compact $AB$ factors derived from low-rank decomposition, thereby eliminating inter-cluster synchronization and achieving near-real-time freshness.
However, \sysname's design faces two critical challenges. \emph{First, the intrinsic low-rankness of updates is dynamic.} 
A fixed LoRA rank is either too small, thus failing to capture complex updates and degrading accuracy, or too large, thus wasting precious memory and compute resources on trivial updates.
\emph{Second, merging training and inference on the same node exacerbates memory bandwidth contention.} 
The irregular access patterns of online training can directly interfere with inference, causing latency spikes that breach the stringent tail-latency SLAs (e.g., P99 latency $<$20ms) and harm QoS.

To address these challenges, \sysname{} introduces two core innovations. First, a \textbf{dynamic rank adaptation} mechanism continuously adjusts the LoRA dimension via real-time PCA monitoring~\cite{hotelling1933analysis} and prunes inactive parameters, effectively constraining memory overhead to less than 2\% of the full EMTs (Section~\ref{sec:memory-management}). 
Second, a \textbf{performance-isolated training runtime} mitigates interference with inference. Specifically, it orchestrates memory access through NUMA-aware resource scheduling and hardware-enforced QoS partitioning. This spatial separation is reinforced by data reuse and load-sensitive scheduling, which jointly reduce contention and minimize interference with critical serving paths. (Section~\ref{sec:training_management}).
Comprehensive evaluation on production-scale datasets demonstrates that \sysname{} effectively resolves the freshness-accuracy tradeoff. Specifically, compared to state-of-the-art delta-update approaches~\cite{matam2024quickupdate}, \sysname{} reduces the update latency by 2×. Furthermore, it achieves higher accuracy than real-time update baselines within tight freshness windows, demonstrating its ability to maintain superior recommendation quality while ensuring near-real-time updates.

In summary, this paper introduces \sysname{}, the first system that fundamentally re-architects DLRM serving by enabling real-time model updates within inference clusters. Our work makes the following key contributions:
\begin{itemize}[leftmargin=*]
    \item \sysname{} achieves sub-second update latency via architectural co-design. By introducing a new paradigm that moves training into inference nodes, \sysname{} reduces DLRM update overhead from over 20 minutes to sub-second, achieving near-real-time freshness.
\item Unlike prior bandwidth-constrained methods that drop critical updates, our low-rank approach captures the full semantics of model changes. \sysname{} outperforms state-of-the-art delta-update baselines by 0.04–0.24\% in accuracy, resolving the fundamental freshness-accuracy tradeoff.
\item We demonstrate the effectiveness and practicality of \sysname{} through a comprehensive evaluation on production-scale traces, showcasing robust performance under real-world conditions.
%
\end{itemize}
\vspace*{-1ex}
\section{Background}
\label{sec:background}  

\subsection{Deep Learning Recommendation Model (DLRM)}
\begin{figure}[t]
  \centering
  \includegraphics[width=0.8\linewidth]{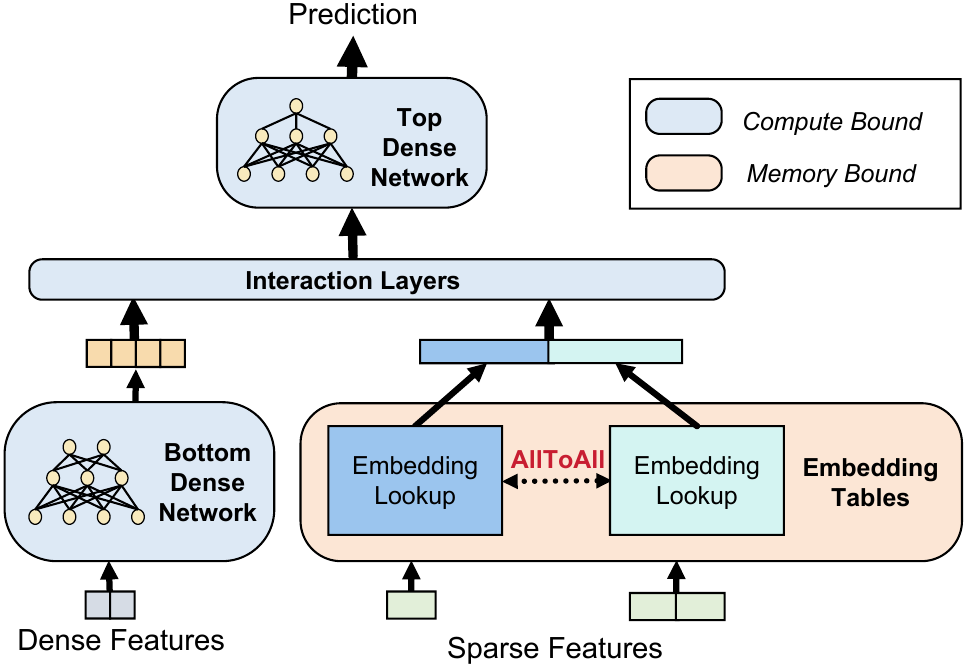}
   \vspace{-1ex}
  \caption{Typical Structure of the DLRM Model}
  \label{fig:dlrm_architecture}
\end{figure}

Deep Learning Recommendation Models (DLRMs) are a class of neural networks pioneered by Meta~\cite{naumov2019deep} for personalized recommendation, widely adopted in industry-scale systems~\cite{naumov2020deep, mudigere2022software, yang2025gpu}. As illustrated in \autoref{fig:dlrm_architecture}, DLRMs process two distinct input types through parallel pathways: 
1) {Dense features}, namely continuous attributes such as user age, are fed into densely connected deep learning networks (e.g., MLPs or transformers~\cite{zhou2018deep, sun2019bert4rec}) to generate dense feature representations. 
2) {Sparse features}, namely categorical attributes, are encoded as one-hot or multi-hot vectors and mapped to dense embeddings via large \emph{embedding tables (EMTs)}. Each EMT corresponds to a categorical field (e.g., gender) and retrieves a unique embedding vector for each categorical ID. For multi-hot inputs, embeddings are pooled (e.g., averaged) to form a single vector.

\textbf{Training} leverages historical interaction data (e.g., user clicks) to optimize parameters end-to-end. Using stochastic gradient descent approaches, the model minimizes loss through forward propagation of training samples and backward updates. Critically, EMTs undergo dynamic row-wise updates: only embeddings corresponding to IDs in each mini-batch are modified, creating irregular memory access. 

During \textbf{inference}, the trained DLRM generates real-time predictions (e.g., click-through probabilities) for incoming requests. The model executes feedforward passes: sparse features trigger embedding lookups from EMTs, while dense features propagate through FC layers. The resulting embeddings and transformed dense features interact via dot products or MLPs to produce prediction scores. 

In industry-scale systems, the size of DLRMs can expand to petabytes.
This massive footprint is primarily attributed to the large EMTs, which store the vast majority of model parameters and are dynamically updated with new interaction data. For instance, Kuaishou, a popular video sharing app, has reported a production system with over 100 trillion parameters, requiring over 200 TB of storage~\cite{lian2022persia}. 
Such scale underscores the critical memory and update challenges inherent to modern recommendation workloads.

\subsection{Industrial DLRM Deployment}
\begin{figure*}[t]
  \centering
  \includegraphics[width=0.65\linewidth]{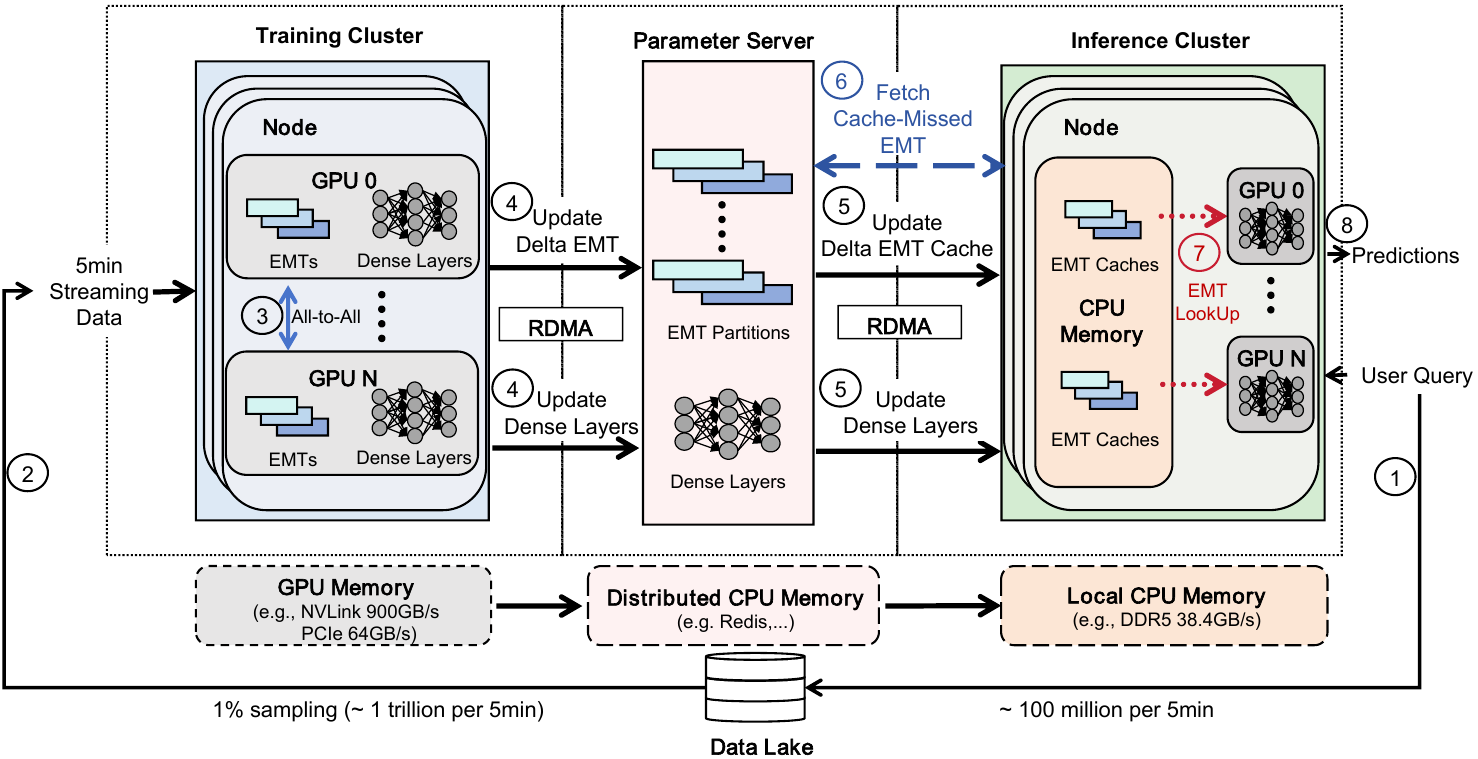}\vspace{-1ex}
  \caption{DLRM Training and Inference System Architecture}
  \label{fig:dlrm_system}
\end{figure*}

Production environments for DLRMs typically employ a decoupled architecture, where training and inference run on separate clusters to meet their distinct computational requirements. 
As shown in \autoref{fig:dlrm_system}, there are three core components in a typical deployed DLRM system, including {training cluster}, {parameter server} and {inference cluster}.

\textbf{End-to-End Workflow.}
The \emph{training cluster} continuously processes streaming user-item interactions to periodically update model parameters, including EMTs for sparse features and dense weights for fully connected layers. 
These updated parameters are then pushed to a central \emph{parameter server}, typically implemented as a sharded key-value store (e.g., Redis), which manages version control and high-throughput writes. The \emph{inference cluster} retrieves the latest parameters from the parameter server to serve real-time recommendations, ensuring consistent model versions across all inference nodes.


\textbf{System Characterization.}
The distinct characteristics of training and inference workloads fundamentally shape the design of production DLRM systems.

\emph{\underline{Training Workload}}.
During training, vast EMTs are partitioned across GPU memory, leveraging all-to-all communications for distributed embedding lookups and gradient synchronization across nodes. This process is demanding due to the following reasons: 1) user requests arrive at extremely high rates (e.g., 30 million queries per 5 minutes in Pinterest's production cluster~\cite{eksombatchai2018pixie}), requiring high batch processing throughput; 2) dynamic, row-wise EMT updates based on mini-batch contents lead to sparse and irregular memory access patterns; and 3) frequent, high-volume parameter exchanges require high bandwidth networks.
To handle these demands, training clusters typically utilize large GPU farms connected via high-bandwidth fabrics (e.g., InfiniBand, NVLink).


\emph{\underline{Inference Workload}}.
In contrast, inference clusters face a different set of challenges: 1) serving latency must be kept within strict bounds (e.g., P99 \textless 20 ms in our recommendations scenario) to ensure real-time responsiveness, which is challenging due to unpredictable request bursts;
and 2) inference clusters must frequently pull fresh parameters from the parameter server to maintain recommendation quality. However, the sheer size of the model parameters means that full synchronization can take minutes, resulting in parameter staleness and degraded serving freshness.

Thus, real production systems typically implement a tiered parameter server architecture which 
1) adopts version batching to group multiple parameter updates into fewer synchronization events, and 2) only transfers delta updates when synchronizing parameter changes. These design choices greatly reduce network load while maintaining model freshness.
The inference cluster employs a hybrid CPU-GPU memory hierarchy, with GPU HBM hosting 5\%-10\% hot embeddings and CPU DRAM (multi-TB per node) storing the remaining warm embeddings~\cite{ren2025machine}. 
By exploiting the inherent skew in embedding access patterns, this hierarchical memory system aims to provide low latency for embedding lookups and massive storage capacity for PB-scale EMTs at the same time.



\subsection{Limitation of Existing DLRMs}

Existing DLRM deployment frameworks face fundamental limitations in supporting timely model updates without compromising accuracy or system stability. 
Production systems like TorchRec~\cite{10.1145/3523227.3547387} and HugeCTR~\cite{wei2022gpu} rely on \emph{full-parameter synchronization}, where entire EMTs are transferred from training clusters to parameter servers during updates.
%
While ensuring strong consistency, this approach proves prohibitively expensive. Synchronizing full parameters (e.g., 200 TB models) over commodity networks (e.g., 100 GbE) require over four hours, introducing unacceptable staleness for real-time recommendations. Moreover, the bursty full-update traffic contends with serving requests, violating stringent P99 latency targets (\textless 20 ms).

To mitigate these limitations, systems like QuickUpdate~\cite{matam2024quickupdate} introduce \emph{delta-based synchronization}. Instead of full EMTs, training clusters select priority parameters to transfer and inference clusters apply partial updates. While reducing transfer costs, this strategy introduces new limitations. First, selection heuristics based on update magnitude (e.g., gradient norms) omit semantically critical but small-value changes (e.g., emerging trends). This could lead to quality degradation issue, which is particularly problematic since even a 0.01\% accuracy loss impacts revenue in production systems~\cite{lin2024understanding}.
Second, to limit quality loss, QuickUpdate recommends 5\% sampling rates for parameters, still yield massive deltas (e.g., 10 TB for 200 TB models). Transferring these deltas consumes \textgreater 14 minutes over 100 GbE network, causing unacceptable delay to model updates.

This landscape reveals a critical research gap: \emph{no existing solution simultaneously achieves low-latency updates and uncompromised inference accuracy}. 
In large-scale production DLRM systems, effective streaming updates must satisfy three core requirements:
\begin{itemize}[leftmargin=*]
    \item \textbf{Accuracy:} Revenue-critical systems tolerate $\leq$ 0.01\% accuracy loss~\cite{lin2024understanding}, while industry studies confirm 5-minute update delays cause measurable performance deterioration~\cite{ktena2019addressing}. This necessitates update mechanisms that maintain near-real-time model freshness to preserve accuracy.
    \item \textbf{Latency:} Industrial DLRM systems enforce strict P99 latency requirements. For example, Meta targets P99 latencies in the tens of milliseconds~\cite{ardestani2022supporting}, while Alibaba mandates serving latencies below 25 ms~\cite{yang2025gpu}.
    \item \textbf{Consistency:} The system must guarantee replica consistency across distributed inference nodes, ensuring identical outputs for the same inputs~\cite{wei2022gpu, yang2025gpu}.
\end{itemize}

\section{Challenges and Motivation}
\label{sec:motivation}  
Existing DLRM systems face critical trade-offs: full-parameter model update ensures consistency but sacrifices model freshness, while delta-update improves latency at the cost of accuracy. To overcome these limitations, \emph{this paper aims to propose a model update mechanism for DLRM systems with both low update cost and high inference accuracy. }

\subsection{Challenges}

\begin{figure}[t]
  \centering
  \begin{subfigure}[t]{0.48\linewidth}
    \centering
    \includegraphics[width=\linewidth]{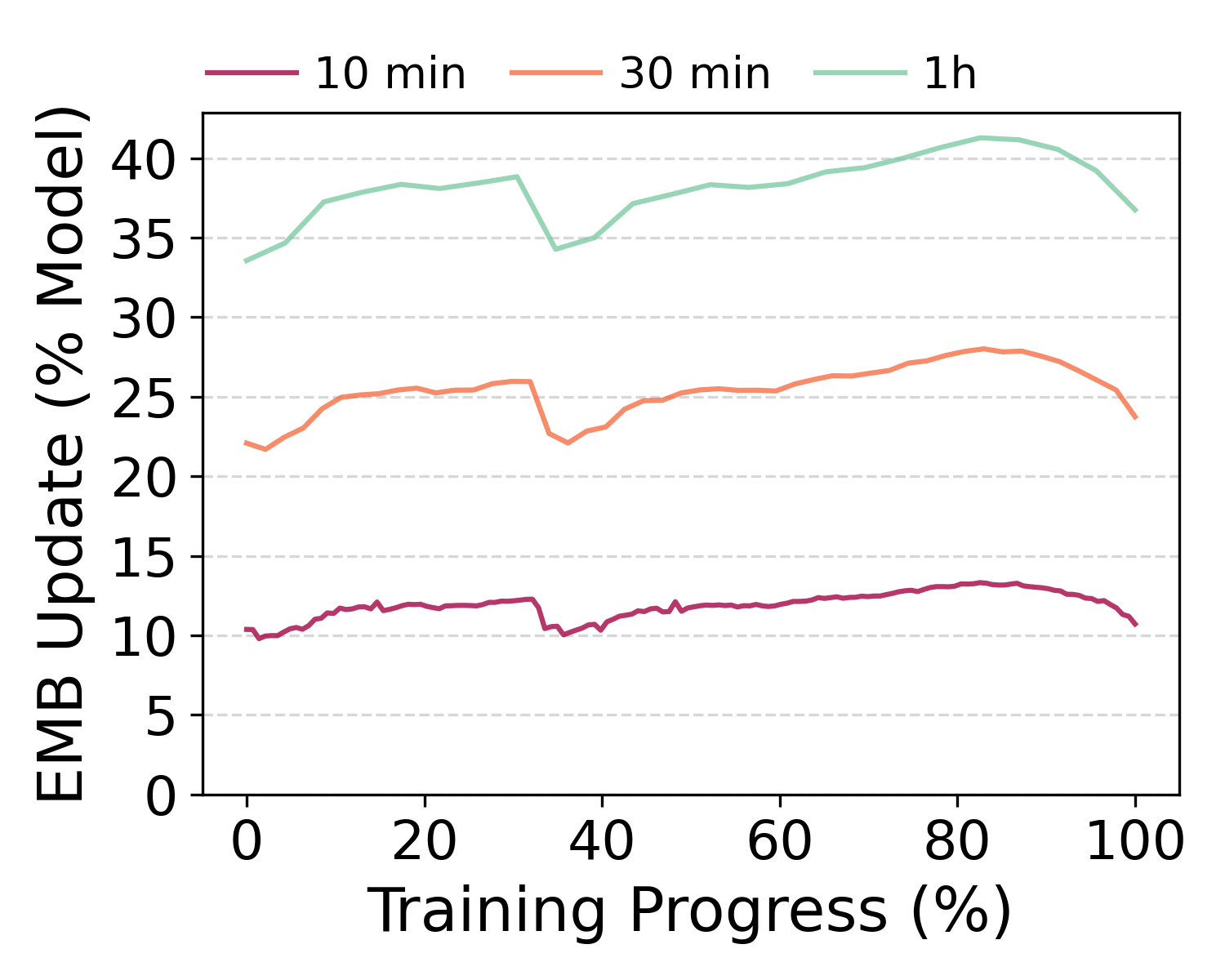}\vspace{-1ex}
    \caption{Embedding update ratio}
    \label{fig:criteo_tb}
  \end{subfigure}
  \begin{subfigure}[t]{0.5\linewidth}
    \centering
    \includegraphics[width=1\linewidth]{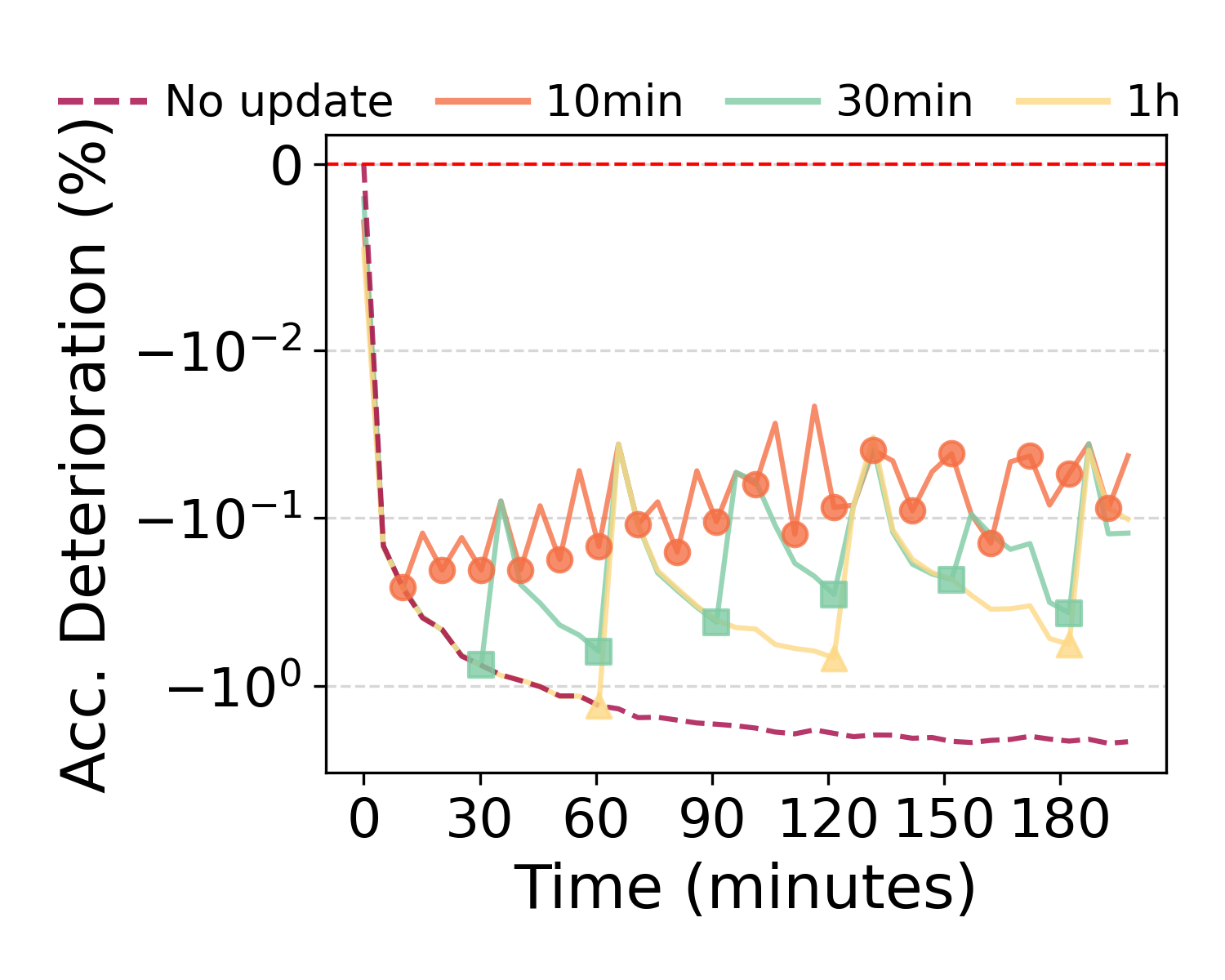}\vspace{-1ex}
    \caption{Accuracy change along serving}
    \label{fig:freq_acc}
  \end{subfigure}
  \vspace{-1ex}
  \caption{Embedding update challenges: (a) model update ratio is high; (b) preserving accuracy requires frequent model updates.}
\end{figure}

Realizing the above objective requires addressing two main challenges inherent to production-scale DLRM systems. 

\textbf{C1: Prohibitive model update volumes.}
Modern DLRMs operate at unprecedented scales, with models like Meta's Model-F~\cite{mudigere2022software} reaching 96TB and Kuaishou's Persia~\cite{lian2022persia} exceeding 200TB for 100 trillion parameters. The primary storage footprint resides in EMTs, which undergo continuous updates as new user-item interactions stream in. 
\autoref{fig:criteo_tb} presents the percentage of embedding parameters updated during model training over 10/30/60-minute intervals, using a production DLRM system in Bytedance. It shows that even with short intervals (e.g., 10-minute), EMTs undergo substantial modifications (e.g., \textgreater 10\%) throughout the training. This persistent high-volume update ratio highlights the continuous synchronization burden in production systems, leading to unacceptable synchronization delay. For example, transferring the 10\% model update may require more time than the available 10-minute update window.
This challenge is further compounded by network contention between update and serving traffic, as well as the need to maintain consistency across distributed parameter replicas.







\textbf{C2: Accuracy degradation under delayed updates.}
Model freshness is crucial for maintaining high recommendation quality. 
To quantify the critical relationship between model freshness and recommendation quality, we study the recommendation accuracy deterioration under different update frequencies using a TB-scale trace from ByteDance (dataset details in Section~\ref{sec:evaluation}).
\autoref{fig:freq_acc} demonstrates the temporal degradation of accuracy due to model staleness during serving. 
As inference progresses without updates, the accuracy declines quickly since outdated embedding parameters fail to capture evolving user-item interactions. Upon model update, the accuracy sharply recovers. 
These observations validate the necessity of high-frequency updates to preserve high accuracy.
{However, this requirement directly conflicts with challenge C1}: increasing update frequency also increases network traffic, leading to higher update cost. This presents a fundamental dilemma: \emph{how can systems deliver frequent updates for freshness without overwhelming network resources?}





\vspace*{-1ex}
\subsection{Opportunities}
\label{sec:design_motivation}
The above challenges necessitate a communication-efficient update strategy that fundamentally \emph{decouples model accuracy from synchronization cost}. 
Through empirical analysis of production DLRM workloads, we identify two intrinsic properties that offer opportunities for realizing this goal.

\begin{figure}[t]
  \centering
  \begin{minipage}[t]{0.48\linewidth}
    \centering
    \includegraphics[width=\linewidth]{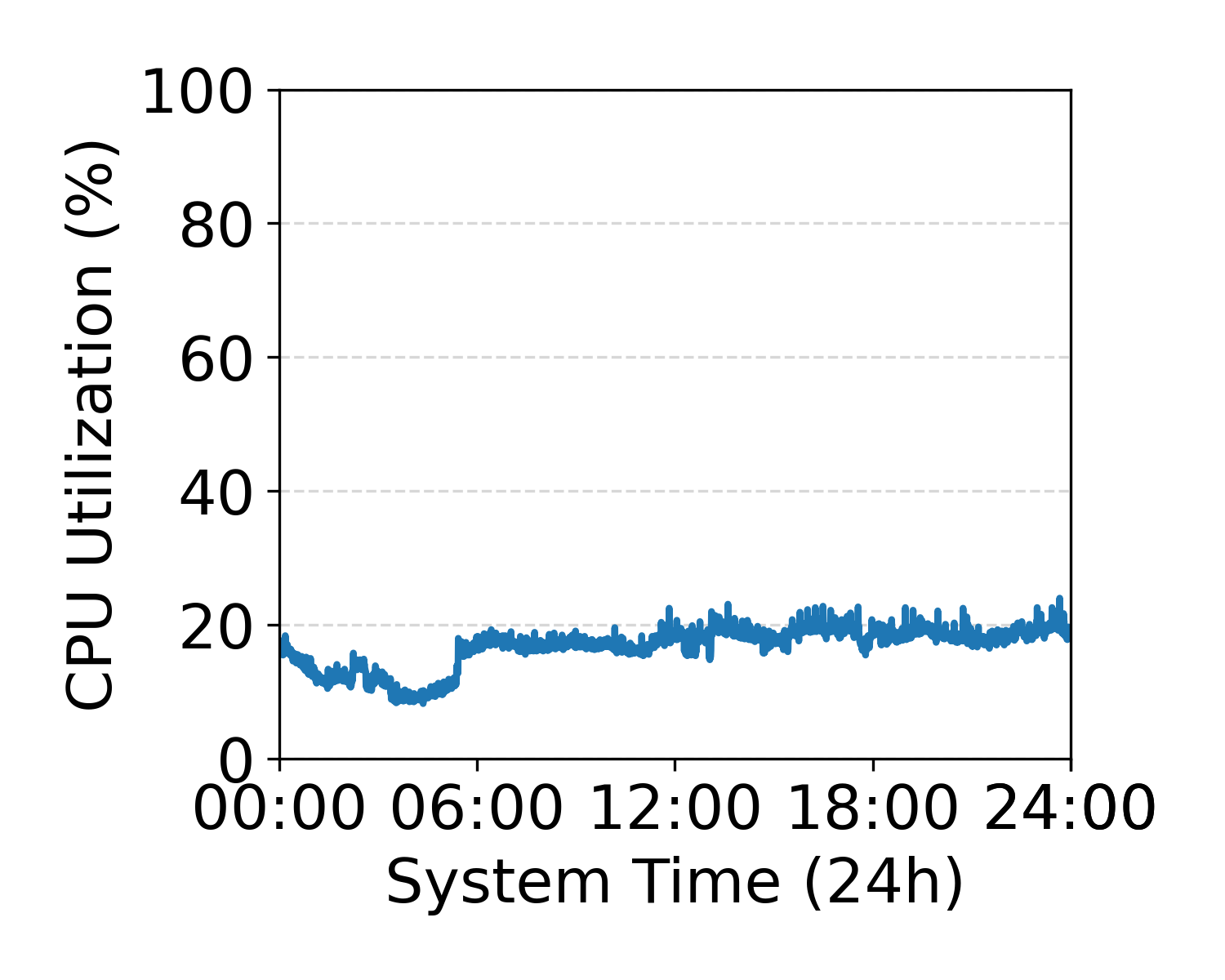}\vspace{-1ex}
    \caption{CPU utilization over 24 hours in ByteDance's production 
    inference cluster}
    \label{fig:cpu_utilization}
  \end{minipage}
  \begin{minipage}[t]{0.48\linewidth}
    \centering
    \includegraphics[width=\linewidth]{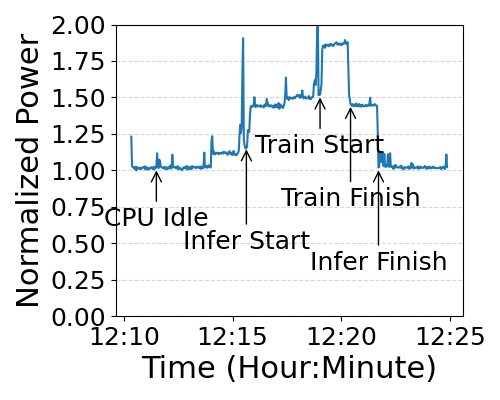}\vspace{-1ex}
    \caption{CPU power variation over 15min in ByteDance's production 
    inference cluster}
    \label{fig:energy-compare}
  \end{minipage}
\end{figure}

\textbf{O1: Underutilized CPUs in inference clusters enable local training.}
Production DLRM deployments utilize a hybrid CPU-GPU architecture in inference clusters, where GPUs accelerate compute-intensive operations (e.g., dense layers) during inference and CPUs host massive EMTs with multi-TB DRAM per node.
Despite substantial memory capacity, CPU resources in ByteDance's inference clusters are severely underutilized. 
As shown in~\autoref{fig:cpu_utilization}, the CPU utilization stays low throughout the day, peaking at only around 20\%. 
Ant Group's trace~\cite{Wang2024DLRover-RM} shows a similar pattern, with CPUs idle for \(\sim\)90\%of the time. 
This creates a significant opportunity: \emph{Idle CPU resources can perform local model updates} to eliminate network transfers and bypass congestion in the already saturated inter-cluster links. 
However, naive implementation where CPUs maintain full parameter replicas for training introduces severe overheads. First, storing duplicate EMTs requires a large memory capacity (e.g., 200TB). Second, full-parameter sync between training/inference states leads to severe network contention for serving, causing a violation of the strict P99 latency requirement. This necessitates lightweight update mechanisms that avoid full replication while maintaining accuracy.



\begin{figure}[t]
  \centering
  \begin{subfigure}[t]{0.49\linewidth}
    \centering
    \includegraphics[width=\linewidth]{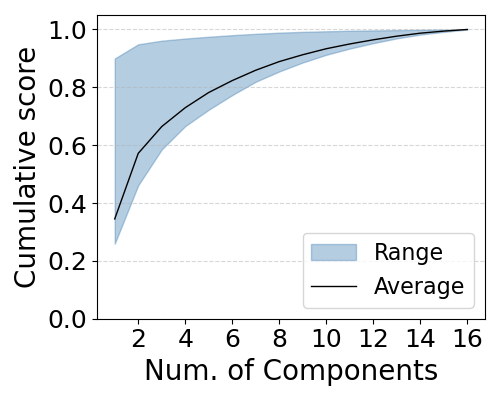}
    \caption{Table with the smallest component spread range}
    \label{fig:pca_table0}
  \end{subfigure}
  \hfill
  \begin{subfigure}[t]{0.49\linewidth}
    \centering
    \includegraphics[width=\linewidth]{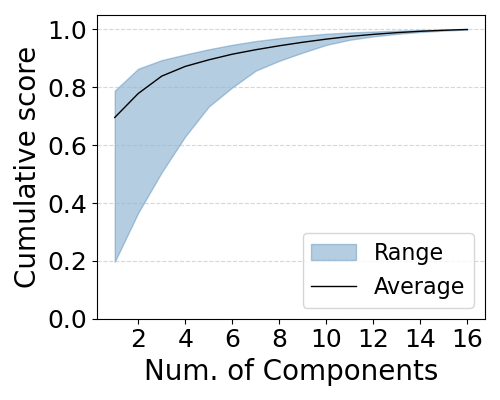}
    \caption{Table with the largest component spread range}
    \label{fig:pca_table1}
  \end{subfigure}
  \caption{Cumulative importance score of PCA components over iterations on two tables in the Criteo dataset~\cite{criteochallenge}. 
  }
  \label{fig:pca_tables}
\end{figure}

\textbf{O2: Low-Rank structure of embedding updates enables efficient training.}
Although EMTs are the primary update bottleneck due to their high-dimensional gradient computations and frequent parameter changes, our key discovery reveals that \emph{the embedding updates ($\Delta W$) exhibit intrinsic low-rank structure}.
Denote an EMT as \( W \in \mathbb{R}^{\mathrm{|V|}\times d} \), where \(\mathrm{|V|}\) is the table size and \(d\) is the dimension of embedding vectors. 
\xg{The low-rank property allows for an efficient, compact approximation of the updates ($\Delta W \approx \Delta W_k$), which is essential for enabling localized training within inference clusters.}

\begin{table}[t]
\centering
\caption{Key notations for low-rank gradient analysis.}
{
\begin{tabular}{ll}
\toprule
\textbf{Symbol} & \textbf{Meaning} \\
\midrule
$W$ & Embedding weight matrix \\
$\Delta W$ & Embedding updates \\
$|\mathrm{V}|$ & Embedding table size (number of entries) \\
$d$ & Embedding dimension \\
$G = \nabla^{(t)}_W \in \mathbb{R}^{|V|\times d}$ & Gradient matrix at iteration $t$\\
$\sigma_i$ & $i$-th singular value of $G$ \\
$G_k = \sum_{i=1}^{k} \sigma_i u_i v_i^{\top}$ & Rank-$k$ approximation of $G$ based on its SVD \\
\bottomrule
\end{tabular}}
\label{tab:notations}
\end{table}

\underline{\emph{Low-rank identification}}.
\xg{The low-rank structure of embedding updates is a fundamental property of over-parameterized models, rigorously demonstrated in existing large language model~\cite{hu2022lora} and DLRM~\cite{huang2020embedding} studies. High-dimensional embedding spaces are intrinsically over-parameterized, while real-world interactions exhibit sparsity, causing gradient directions cluster in low-dimensional subspaces~\cite{huang2020embedding}.}\label{re:Q5}

\xg{To validate this, we perform Principal Component Analysis (PCA) on the gradient matrix, $G = \nabla_W^{(t)} \in \mathbb{R}^{|V| \times d}$, across multiple training iterations $t$.
Figure~\ref{fig:pca_tables} presents the cumulative variance explained by the top-$k$ principal components for two representative EMTs from the Criteo dataset~\cite{criteochallenge}. 
Our observation is that, only a few principal components capture the majority of the variance in $G$.
For instance, in Figure~\ref{fig:pca_table1}, the top 3 components capture 80\% of the gradient information in the best case, and in the worst case (Figure~\ref{fig:pca_table0}), only 6 components are sufficient to cover 80\% of the variance.
Similar patterns have also been observed using other production and benchmark datasets in Table~\ref{tab:datasets_all}.
These results reveal updates to the EMTs primarily lie along few dominant directions, thus confirming the intrinsic low-rank structure of the gradients.
}

\underline{\emph{Low-rank solution}}.
\xg{The empirical evidence motivates a Low-Rank Adaptation (LoRA) representation for the updates.
To determine the appropriate compact rank $k$, we use the statistical properties revealed by PCA.

Formally, the gradient matrix $G$ is approximated using the Eckart–Young–Mirsky theorem [12], which provides the optimal rank-$k$ approximation ($\hat{G}_k$) under the Frobenius norm:
\begin{equation}\label{eq:svd}
    G \approx \hat{G}_k = \sum_{i=1}^{k} \sigma_i u_i v_i^\top = U_k \Sigma_k V_k^\top 
\end{equation}
where $\sigma_i$ are the singular values. 

The rank $k$ is chosen as the smallest value that preserves a target fraction ($\alpha$) of the total gradient variance:
\begin{equation}
    \label{eq:variance_threshold}
    \frac{\sum_{i=1}^k \sigma_i^2}{\sum_{j=1}^d \sigma_j^2} \geq \alpha
\end{equation}

This criterion allows the system to dynamically determine the minimum necessary rank ($k$) to meet a desired accuracy target. Based on empirical findings, we use the threshold $\alpha \in [{0.8},{0.95}]$ to strike an effective balance between accuracy and memory efficiency. By default, $\alpha$ is set to 0.8.
Since $\Delta W$ is accumulated from these low-rank gradients, the overall update matrix also exhibits low-rank characteristics, fully justifying our use of LoRA modules for efficient synchronization.

}

\subsection{Design Motivation}



In summary, the dual challenges of prohibitive update volumes (C1) and accuracy degradation under delayed updates (C2) create an intractable tension for production DLRM systems. 
Our analysis of production environments reveals two key opportunities to resolve this tension: Inference clusters contain substantial underutilized CPU capacity (O1), while embedding updates exhibit intrinsic low-rank structure (O2). Together, these insights enable a novel approach named \textbf{\sysname{}}, which \textbf{leverages idle inference-cluster CPUs to compute and apply low-rank embedding updates locally.}

This design possess several key advantages:
\begin{itemize}[leftmargin=*]
    \item By performing lightweight model update using low-rank approximations on inference nodes, we avoid inter-cluster synchronization entirely. This significantly reduces update latency to maintain model freshness.
    \item Through variance-controlled low-rank approximation (e.g., $\alpha \geq 0.8$), \sysname{} guarantees a theoretically-bounded accuracy loss compared to full-parameter synchronization.
    \item The above two benefits are achieved with minimal cost. As measured in production environment (\autoref{fig:energy-compare}), the CPU power of running concurrent inference and training is only 20\% higher than inference-only operations. This cost is marginal compared to the achieved memory compression and communication reduction.
\end{itemize}

\section{Design Details}
\label{sec:proposed_method}  


\begin{figure*}[t]
  \centering
  \includegraphics[width=0.8\linewidth]{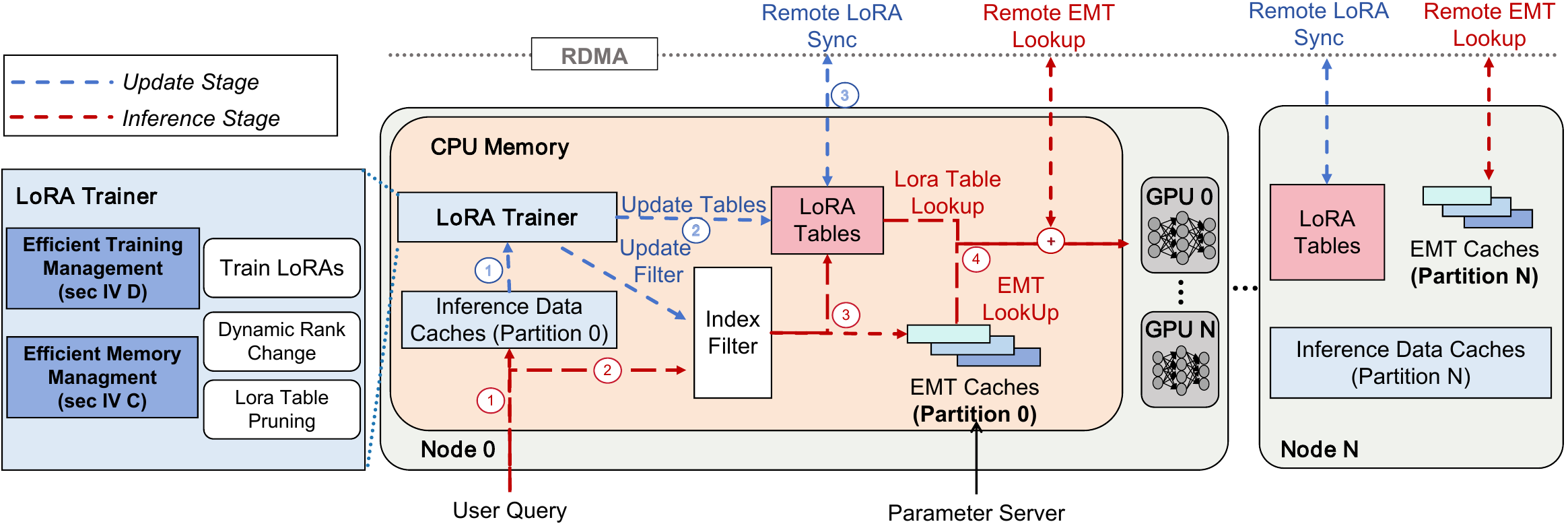}
  \caption{Overall architecture of \sysname{}}
  \label{fig:proposed_architecture}
\end{figure*}

In this section, we introduce the design considerations and proposed optimizations in \sysname{}.

\subsection{System Overview}

\autoref{fig:proposed_architecture} shows the overall architecture of \sysname{}, which integrates online embedding update directly within the inference workflow.
Specifically, as inspired by the low-rank solution discussed in ~\ref{sec:design_motivation}, we adopt Low-Rank Adaptation (LoRA)~\cite{hu2022lora}, a parameter-efficient fine-tuning method, to decompose model updates into low-rank matrices: 
\begin{equation}\label{eq:lora}
\Delta W = AB, \quad A \in \mathbb{R}^{|V| \times k}, \quad B \in \mathbb{R}^{k \times d}
\end{equation}
where $k\ll d$ (e.g., 3-6 vs. 16 in \autoref{fig:pca_tables}).
Mathematically, this is equivalent to the optimal low-rank approximation of $\Delta W$ derived via truncated SVD. This decomposition reduces the update payload size by a factor of $d/k$ (e.g., 5x), making in-node computation and storage feasible.

As shown in the figure, \sysname{} operates through two concurrent pipelines: a low-latency inference path and a resource-aware online update path, which share critical state through carefully designed in-memory data structures.
Below we briefly introduce the data path in the two pipelines.

\textbf{Inference Path (Red).} The primary function of the node is to serve inference queries satisfying strict latency constraints. 
\textcircled{1} A user query enters the system, triggering feature extraction.
\textcircled{2} The Hot Index Filter checks if each sparse feature ID has been recently updated by the update path. 
\textcircled{3} For a ``hot'' ID, the embedding is computed as $W_\text{base}[i] + A[i]B$. Both the base embedding $W_\text{base}[i]$ and the LoRA factors $A[i]$ are fetched from the local LoRA Cache (hosted in CPU DRAM).
For a ``cold'' ID, only the base embedding $W_\text{base}[i]$ is retrieved from the local cache or, on a miss, from a remote parameter server.
\textcircled{4} The processed batch of embeddings is concatenated and sent to the GPU for forward propagation through the dense layers of DLRM to generate a prediction.
Crucially, the original query and feature IDs are cached into a shared data buffer in CPU DRAM. The buffer serves as the training dataset for the online update path.

\textbf{Online Update Path (Blue).} This path leverages idle CPU cores to continuously refresh the model from cached inference data.
\textcircled{1}
\emph{Training}: At a fixed interval, a training thread samples a mini-batch of feature-index pairs (usually a five-minute time window) from the shared buffer.
It performs a forward and backward pass. Gradients are computed only for the low-rank matrices $A$ and $B$, while the base weights $W$ remain frozen.
Before updating $A$ and $B$ using computed gradients, our dynamic rank and pruning mechanisms (Section~\ref{sec:memory-management}) adjust the size of $A$ and $B$ and evict infrequently updated rows to control memory footprint.
\textcircled{2} \emph{Update}:
The updated LoRA factors are stored in the local LoRA table cache, driving the next round of inference. 
\textcircled{3} \emph{Sync}: To maintain consistency across inference nodes, updated rows of $A$ are asynchronously synced with other nodes via a non-blocking AllGather operation every $T$ steps. This design provides eventual consistency, prioritizing inference availability over strict instantaneous coherence, a practical trade-off for recommendation systems. 


\textbf{Design Challenges}.
\sysname's design faces two critical challenges: 1) \emph{Dynamic Rank Instability}: The intrinsic low-rankness of embedding updates is highly variable, as shown in ~\autoref{fig:pca_tables}. A static LoRA rank 
$k$ is inherently inefficient, leading to either accuracy degradation (rank too low) or resource waste on negligible updates (rank too high). This necessitates a runtime mechanism to adapt 
$k$ in real-time.
2) \emph{Memory Bandwidth Contention}: Co-location of training and inference risks severe resource conflict. The random, write-intensive memory access of the LoRA trainer directly contend with the read-intensive inference workload. This contention for shared DRAM bandwidth dramatically increases latency, causing P99 latency violations and degrading user experience.

The following subsections present our solutions to these challenges. Section~\ref{sec:memory-management} introduces a dynamic adaptation mechanism to solve the instability of rank and memory size. Section~\ref{sec:training_management} details our resource management system to eliminate bandwidth contention and ensure performance isolation.

\begin{figure*}[t]
  \centering
  \includegraphics[width=0.8\linewidth]{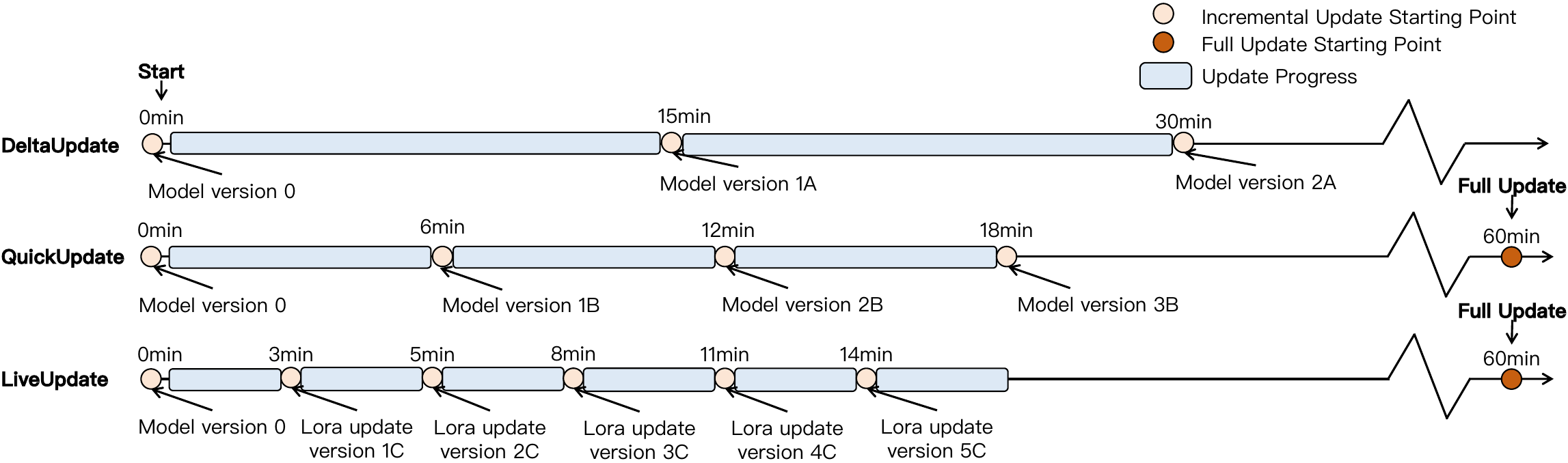}
  \caption{\xg{Model update timeline of three compared methods. All systems start from identical model version 0, and apply different update strategies. QuickUpdate and \sysname{} periodically (e.g., hourly) perform full-parameter update to limit model drift. }}
  \label{fig:time_line}
\end{figure*}

\begin{figure}[t]
  \centering
  \begin{minipage}[t]{0.49\linewidth}
    \centering
    \includegraphics[width=\linewidth]{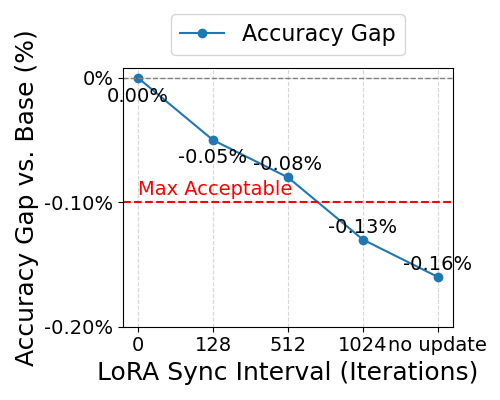}
    \caption{Accuracy gap under different LoRA sync intervals.}
    \label{fig:sync_freq}
  \end{minipage}
  \hfill
  \begin{minipage}[t]{0.49\linewidth}
    \centering
    \includegraphics[width=\linewidth]{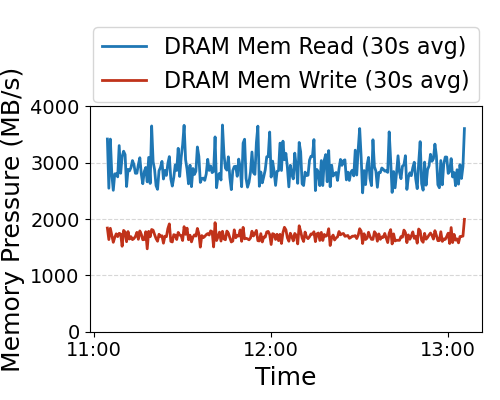}
    \caption{The DDR's memory pressure during Inference.}
    \label{fig:mem-pressure}
  \end{minipage}
\end{figure}
\subsection{Online Model Update Workflow}
\autoref{fig:time_line} compares the model update timelines of \sysname{} with two existing approaches, namely DeltaUpdate (industry practice, see Section~\ref{sec:eval:config}) and QuickUpdate~\cite{matam2024quickupdate} (current SOTA). 
All methods start from the same initial model (Version 0) and apply different update strategies, resulting in different model evolution over time. 
In particular, \sysname{} follows a tiered update strategy: 1) At short-term (e.g., 0-60min), it applies \emph{local} low-rank adaptations using idle inference resources; 2) At mid-term (e.g., per hour), it triggers full-parameter synchronization using updates pulled from the training cluster to prevent \emph{model drift accumulation}; 3) At long-term (days/weeks), full retraining could happen in case of fundamental model updates (e.g., new architecture).

The figure illustrates that \sysname{} could perform the most frequent updates due to its low update overhead, delivering better model freshness than the other two approaches. Similar to \sysname{}, QuickUpdate also adopts hourly full-parameter synchronization to limit model drift~\cite{matam2024quickupdate}. 

\subsection{Dynamic Rank Adaptation for Memory Efficiency}
\label{sec:memory-management}  
The LoRA table stores the $A$ matrices for all active embedding indices. For each active embedding index $i$, it stores a row $A[i]\in \mathbb{R}^{1 \times k}$. 
According to equation~\ref{eq:lora}, the LoRA table size is determined by $|V|$ and low-rank dimension \( k \). The compressed table size is about \(k/d\) of the original EMT (e.g., 3/16 in \autoref{fig:pca_tables}), which is still too large for practical deployment.
To mitigate this issue, we introduce two complementary lightweight adaptation techniques to minimize memory overhead.
These techniques dynamically adjust the LoRA table size and rank based on the current training workload, ensuring that only the most relevant indices are retained while minimizing memory usage.

\textbf{Variance-Aware Rank Adaptation.}
The intrinsic dimensionality (the optimal $k$) of gradient updates is not static but evolves during training. A fixed rank is therefore inefficient. 
To optimize LoRA size while preserving model accuracy, we propose to adaptively adjust LoRA rank \( r \) based on the recent gradient updates. We observe through \autoref{fig:pca_tables} that model updates exhibit a \textbf{dynamic} low-rank 
structure that evolves during training. Thus, we periodically calculate the LoRA rank at each fixed interval \( T \) (e.g., every 128 iterations).

Our solution is to dynamically find the minimal rank that preserves the most ``important'' information in the updates. We treat the recent gradient updates as a set of vectors and use PCA to identify the dominant directions of change. Concretely, at a fixed interval $T$, we perform PCA on a snapshot of these gradients. The eigenvalues ($\lambda_j$) from PCA represent the importance of each new direction. We then find the smallest rank $r_t$ such that the sum of the top-$r_t$ eigenvalues captures a large fraction $\alpha$ (e.g., 80\%) of the total variance. The new global rank $r$ is then set to the average of these observed ranks over the interval to smooth out transient fluctuations:
\[
r = \left\lceil \frac{1}{T} \sum_{t=1}^{T} r_t \right\rceil, \quad \text{where } r_t = \arg\min_{r'} \left( \frac{\sum_{j=1}^{r'} \lambda_j}{\sum_{j=1}^{d} \lambda_j} \geq \alpha \right)
\]

The new rank $r$ is used to resize the LoRA matrices, ensuring they are always optimally sized to capture the essential information and avoid both under- and over-compression.

\begin{figure}[t]
  \centering
  \includegraphics[width=0.8\linewidth]{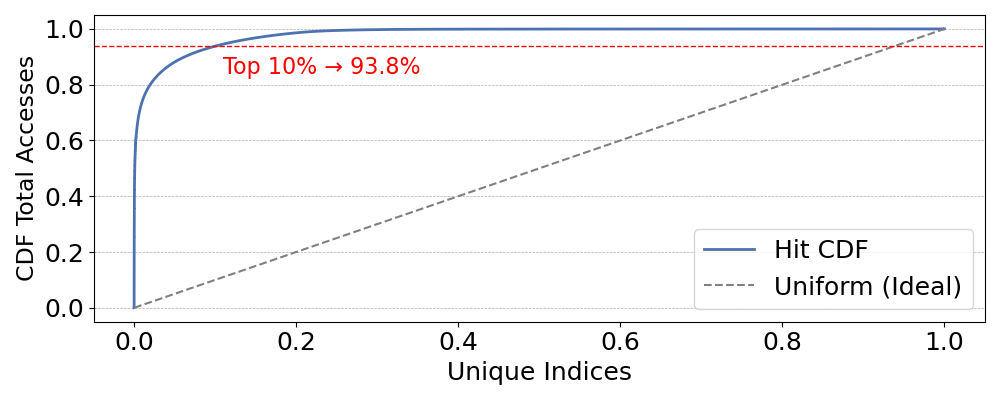}\vspace{-1ex}
  \caption{CDF of embedding access distribution.}
  \label{fig:hotspot_dist}
\end{figure}

\textbf{Usage-Based Table Pruning.}
While the rank $k$ controls the width of the LoRA table, we must also manage its length. A naive implementation allocates a row for every embedding index. However, we observe that the majority indices are updated infrequently, causing memory waste.
We track the update frequency \( f_i \) of each embedding index \( i \) over a sliding window of \( T \) iterations. If an index has not been updated for a certain threshold \( \tau_{\mathrm{prune}} \), it is considered inactive and can be pruned. We define the index that updated more than \( \tau_{\mathrm{prune}} \) times in the last \( T \) iterations as an active index, and the set of active indices is denoted as \( \mathcal{I}_{\mathrm{active}} \). The active indices are used to determine the current size of the LoRA table \( C_t \) at each interval \( t \). We adjust the LoRA table size dynamically according to:
\begin{equation} 
  C_{t+1} = \min\left( \max\left( |\mathcal{I}_{\text{active}}|, C_{\min} \right), C_{\max} \right) 
\end{equation}
where \( C_{t+1} \) is the size of the LoRA table in the next interval, \( \mathcal{I}_{\mathrm{active}} \) is the set of active indices that have been updated more than \( \tau_{\mathrm{prune}} \) times in the last \( T \) iterations. \( C_{\min} \) is the minimum allowable size of the table (defaulting to \( \frac{1}{50} \) of the full table size), and \( C_{\max} \) is the full size of the embedding table.
\begin{algorithm}[t]
  \footnotesize
  \caption{Adaptive LoRA Memory Management}
  \label{alg:adaptive_lora}
  \begin{algorithmic}[1]
    \State \textbf{Input:} Recent gradients $G$, variance threshold $\alpha$, pruning threshold $\tau_{\text{prune}}$, size bounds $[C_{\text{min}}, C_{\text{max}}]$
    \State \textbf{Output:} Updated rank $r$, updated active set $\mathcal{I}_{\text{active}}$
    
    \State $r \gets$ \textproc{Rank\_Adaptation}($G$, $\alpha$); \Comment{Adapt LoRA rank using PCA}
    \State $\text{Resize } A^{(t)}, B^{(t)} \text{ to rank } r;$ \Comment{Reconfigure LoRA factors}
    \State $\mathcal{I}_{\text{active}} \gets \emptyset$;  \Comment{set $\mathcal{I}_{\text{active}}$ into empty set}
    \For{each index $i$ in table}
        \If{$f_i \geq \tau_{\text{prune}}$} \Comment{$f_i$ is update freq. in last $T$ steps}
            \State $\mathcal{I}_{\text{active}} \gets \mathcal{I}_{\text{active}} \cup \{i\}$;
        \EndIf
    \EndFor
        \State $C_{\text{new}} \gets \min(\max(|\mathcal{I}_{\text{active}}|, C_{\text{min}}), C_{\text{max}})$; \Comment{Clamp the new size}
    \State $\text{Resize table to } C_{\text{new}}; \text{ initialize new rows to } 0$;
    \State \Return $r$, $\mathcal{I}_{\text{active}}$ \Comment{Return new configuration}
    
  \end{algorithmic}
\end{algorithm}
\begingroup
\setlength{\parskip}{0.1em} 

To minimize performance impact, our memory management strategies run asynchronously in a background thread and are triggered only every few hundred training steps. This ensures the process is lightweight and does not interfere with the main training loop. Algorithm~\ref{alg:adaptive_lora} implements our adaptive memory management strategy with minimal runtime overhead. Every $T$ iterations, the system performs three key operations: (1) applying \textproc{RankAdaptation} to recent gradient snapshots to determine an optimal LoRA rank $r_t$ (Line 3), (2) reconfiguring the factor matrices $A^{(t)}$ and $B^{(t)}$ with this new rank (Line 4), and (3) pruning the embedding indices based on usage frequency (Line 5-10). By identifying indices with $f_i < \tau_{\mathrm{prune}}$ and removing them from the active set $\mathcal{I}_{\mathrm{active}}$, we maintain only frequently accessed embeddings. The table size is then dynamically adjusted within the bounds $[C_{\min},C_{\max}]$ to accommodate exactly the required number of active indices. 
\xg{\( \tau_{\mathrm{prune}} \) is a parameter in Algorithm~\ref{alg:adaptive_lora}. As shown in \autoref{fig:criteo_tb}, the changed ratio within a 10-minute window is about 10\%, 
and \autoref{fig:hotspot_dist} shows that top 10\% indices account for 93.8\% accesses. 
Hence, we initialize the LoRA table with 10\% of its full size and set the pruning threshold 
\( \tau_{\mathrm{prune}} \) to the access frequency of rank 10\%. \( \tau_{\mathrm{prune}} \) can be dynamically updated by tracking embedding access frequencies to maintain the top-10\% boundary during the inference period. }

Our experiments show that this joint adaptation reduces the memory footprint to about \textbf{2\%} of the original LoRA table while preserving accuracy. For production-scale recommendation systems with multi-terabyte embedding tables, this makes real-time adaptation feasible, turning an otherwise unrealistic memory demand into a deployable solution. Separately, prior work has explored LoRA compression and factorization methods~\cite{Gao2024Parameter-Efficient,Kim2024RA-LoRA}, achieving up to $20\times$ reduction. Such techniques could further improve the scalability of \sysname{}, though they are not the focus of this paper.
\endgroup

\subsection{Performance Isolation via NUMA-Aware Scheduling}
\label{sec:training_management}


\begin{figure}[t]
  \centering
  \begin{subfigure}[t]{0.48\linewidth}
    \centering
    \includegraphics[width=\linewidth]{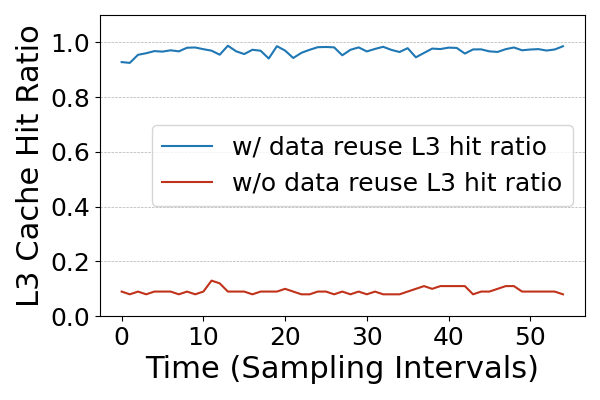}
    \caption{Data Reuse}
    \label{fig:l3_seq_vs_rand}
  \end{subfigure}
  \hfill
  \label{fig:l3cache}
  \begin{subfigure}[t]{0.48\linewidth}
    \centering
    \includegraphics[width=\linewidth]{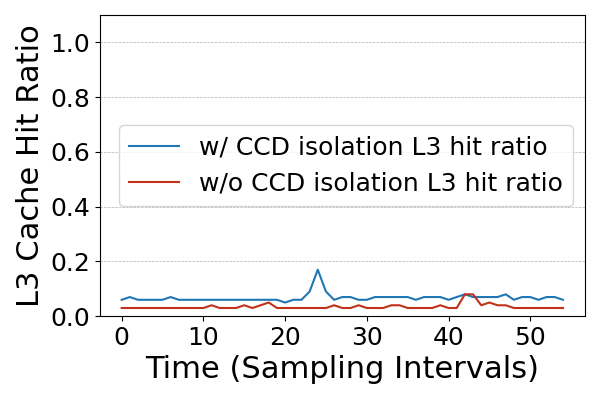}
    \caption{CCD scheduling}
    \label{fig:l3_rand_vs_mix}
  \end{subfigure}\vspace{-1ex}
  \caption{L3 cache hit ratio of the LoRA training process and inference process on a single node, before and after using our (a) data reuse and (b) CCD scheduling optimizations}
\label{fig:l3_hit}
\end{figure}

Co-locating training and inference risks severe performance interference. 
Although memory bandwidth may not be fully saturated (\autoref{fig:mem-pressure}), our experiments demonstrate that simply co-locating training and inference without mitigation increase P99 latency by over $2\times$ (\autoref{fig:latency-ablation}), violating strict service-level agreements. 
This degradation occurs because the LoRA trainer contend with the inference for memory bandwidth.
Our analysis using hardware performance counters (perf~\cite{perf_tools}) confirms that contention drastically increases last-level cache (LLC) miss rates and memory access latency. As shown in \autoref{fig:l3_hit}, the L3 hit rate for both training and inference workloads are less than $<10\%$ (red lines) without optimizations.

To mitigate this interference, we propose two interdependent techniques that leverage hardware awareness and data reuse to eliminate contention.

\textbf{Embedding Vector Reuse.} 
Our first technique exploits the insight that, the inference engine has already fetched and computed necessary embedding vectors in memory. To eliminate redundant and expensive memory lookups, our system allows the training process to reuse the cache-hot embedding vectors from the inference phase, through \emph{a shadow embedding table}. 
We ensure this sharing is efficient through several key design choices. First, vectors are pinned contiguously in DRAM via \texttt{mlock}~\cite{linux_mlock} to eliminate swapping and page fault overhead. Second, NUMA-aware allocation co-locates these locked pages with the training threads to ensure local access. Finally, software prefetching (e.g., via \texttt{\_\_builtin\_prefetch(addr,0,3)})~\cite{gcc_builtin_prefetch} proactively moves data into cache, further increasing hit rates.
Together, these optimizations transform training's memory access patterns into a cache-friendly operation, eliminating swap-induced latency spikes and significantly reducing memory traffic.

\textbf{NUMA-Aware Resource Scheduling.}
To eliminate cache-level interference, we leverage the non-uniform memory access (NUMA) architecture of modern servers. Our approach is motivated by two key observations. First, existing work have revealed that the embedding accesses follow a power-law distribution~\cite{ardestani2022supporting,wei2022gpu}, where over 90\% of requests target less than 10\% of indices. Second, modern CPUs (e.g., AMD EPYC) organize cores into multiple Core Complex Dies (CCDs), each with a private L3 cache large enough (e.g., 96MB) to hold these hot embeddings.

Based on these insights, our strategy is to spatially isolate the workloads onto dedicated CCDs. We pin the latency-critical inference threads to one set of CCDs and the training threads to another. This ensures that the ``hot'' working set of each workload (the frequently accessed embeddings) remains cached within its own local L3, preventing the destructive cross-workload cache thrashing that causes latency spikes. 
To maintain optimal performance under dynamic conditions, we propose an adaptive algorithm (Algorithm~\ref{algo:ccd_alloc}) to continuously monitor latency metrics and dynamically rebalance CCD assignments between inference and training, ensuring strict QoS targets are met.
Specifically, before each training cycle, the system monitors P99 inference latency over a short window (Line 6). If the GPU inference time's latency exceeds a high threshold (e.g., 10ms), one CCD is relocated from training to inference (Line 7-8). If latency drops below a low threshold (e.g., 6\,ms) and training has not reached its cap, one CCD is moved back to training (Line 9-10). All adjustments respect the minimum inference requirement and training cap to avoid saturating memory bandwidth. All adjustments are followed by an immediate rebinding of processes to their new CCD assignments (Lines 11-12). The controller then enters a sleep period until the next cycle (Line 13).

\begin{figure}[t]
  \centering
  \includegraphics[width=0.8\linewidth]{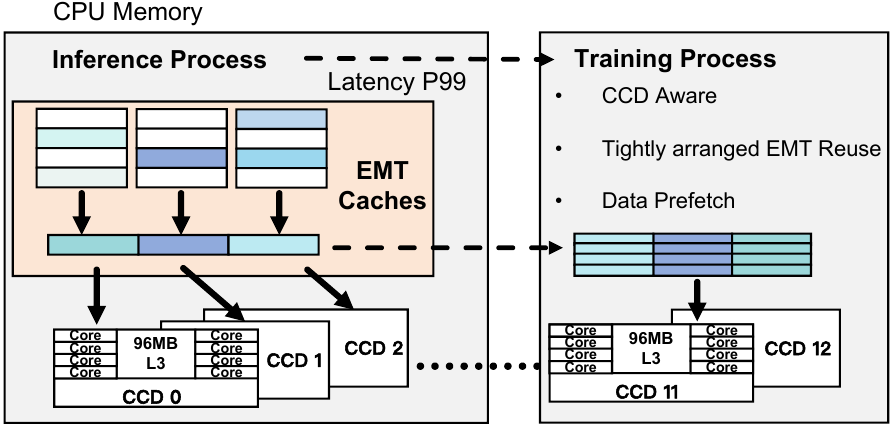}
  \caption{Numa-Aware and Data-Reuse Techniques.}
  \label{fig:cache_accel}
\end{figure}

~\autoref{fig:cache_accel} illustrates an example of the two methods above. Initially, the inference process holds 10 CCDs while the training process holds 2 CCDs. During inference, the process stores the embedding vectors into a tightly arranged shared buffer, allowing the training process to reuse them. Meanwhile, the training process monitors the Latency P99 of the inference process and dynamically adjusts its CCD allocation accordingly.


\xg{Note that, although we use AMD EPYC CCDs to demonstrate NUMA-aware memory optimizations, our algorithm requires no CCD-specific features and maintains equal effectiveness on standard multi-core architectures. Further, the CCD abstraction is widely adopted in modern CPU-based optimizations~\cite{jain2025load,Nair2024Parallelization}, making it a standard practice.
}

\begin{algorithm}[t]
  \footnotesize
  \color{black}
  \caption{Adaptive NUMA Resource Partitioning}
  \label{algo:ccd_alloc}
  \begin{algorithmic}[1]
    \State \textbf{Input:} $C$ = set of all CCDs; $T_{\mathrm{high}}$ = high-latency threshold; $T_{\mathrm{low}}$ = low-latency threshold; $m_{\mathrm{inf}}$ = min inference CCDs; $M_{\mathrm{train}}$ = max training CCDs; $T_{\mathrm{mon}}$ = monitoring window; $T_{\mathrm{cycle}}$ = cycle period.
    \State \textbf{Output:} $C_{\mathrm{inf}}$ = inference CCD set; $C_{\mathrm{train}}$ = training CCD set.
    \State Initialize $C_{\mathrm{inf}}\leftarrow$ any $m_{\mathrm{inf}}$ CCDs \Comment{pick initial inference devices}
    \State $C_{\mathrm{train}}\leftarrow C \setminus C_{\mathrm{inf}}$ \Comment{the rest go to training}
    \For{each adaptation interval $t$}
      \State $p_{99}\leftarrow \mathrm{measure\_p99}(\mathrm{inference}, T_{\mathrm{mon}})$ \Comment{monitor current p99 latency}
      \If{$p_{99}\ge T_{\mathrm{high}}$ \textbf{and} $|C_{\mathrm{inf}}|<|C|-m_{\mathrm{inf}}$}
        \State move one CCD from $C_{\mathrm{train}}$ to $C_{\mathrm{inf}}$ \Comment{add capacity to inference}
      \ElsIf{$p_{99}\le T_{\mathrm{low}}$ \textbf{and} $|C_{\mathrm{train}}|<M_{\mathrm{train}}$}
        \State move one CCD from $C_{\mathrm{inf}}$ to $C_{\mathrm{train}}$ \Comment{reclaim for training}
      \EndIf
      \State \text{rebind}(\text{inference}, $C_{\mathrm{inf}}$) \Comment{apply new inference binding}
      \State \text{rebind}(\text{training},  $C_{\mathrm{train}}$) \Comment{apply new training binding}
      \State sleep($T_{\mathrm{cycle}}$) \Comment{wait until next adjustment}
    \EndFor
  \end{algorithmic}
\end{algorithm}

\subsection{System Implementation}
\label{sec:system_implementation}
Deploying out design in a real-world inference cluster need to solve several 
practical challenges.

\textbf{Training Data from Inference Logs.}
A primary challenge for deploying \sysname{} is the absence of a traditional training pipeline on inference nodes. To generate training data, we cache feature IDs and their associated labels from real-time user requests into a ring buffer with a 10-minute retention window. This provides a continuous stream of fresh, real-world training samples directly sourced from production traffic. The data volume is manageable—approximately 40-50 GB per window—making it feasible to hold in memory or local SSD without contention.

\textbf{Communication-Efficient LoRA Synchronization.}
In standard data-parallel training, each rank maintains a copy of the low-rank adapter $\theta_r$ and updates it using its local mini-batch. Despite LoRA’s small parameter size, naively all-reducing every gradient still incurs unacceptable communication overhead.
To solve this, we introduce a \textit{parallel-update with priority-merge} protocol (Algorithm~\ref{alg:lora_sync}). 

Algorithm~\ref{alg:lora_sync} details the memory-optimized training procedure in four phases. 
(1) Initialization: The system initializes the parameters and assigns an empty local index set $S_r$ to each rank $r$. This set $S_r$ acts as a sparse tracker for modified parameters, eliminating the memory overhead of storing a redundant reference copy $\theta_{\mathrm{last}}$ (a snapshot of the model before updates) (Line 3).
(2) Local Training: Each rank $r$ independently updates its local model parameters $\theta_r$. Crucially, instead of comparing full dense tensors, the system records the \textit{support} of the update—denoted as $\mathrm{supp}(\Delta \theta_r)$, which represents the specific indices where the parameter change is non-zero. These indices are accumulated into the set $S_r$ (Lines 5--7).
(3) Priority Merge: At synchronization intervals, the system computes $\mathcal{I}_{\mathrm{all}}$, which is the global union of all indices modified by any rank. Write conflicts are resolved deterministically using a rank-priority rule: for a specific parameter index $i$, the global model adopts the value from the winning rank $k$. Here, $k$ is defined as the maximum rank ID among all ranks that modified index $i$ (formally, $k = \max \{r \mid i \in S_r\}$) (Lines 8--11).
(4) Broadcast: Finally, the unified updates are broadcast to all ranks, and the tracking sets $S_r$ are reset to empty for the next cycle (Line 12).

\begin{algorithm}[t]
  \footnotesize
  \caption{Sparse Data-Parallel LoRA with Periodic Sync}
  \label{alg:lora_sync}
  \begin{algorithmic}[1]
    \State \textbf{Input:}  Ranks $R$, Interval $T_{\mathrm{sync}}$, Params $\theta$, Learning rate $\eta$
    \State \textbf{Output:}  Optimized parameters $\theta$
    \State Initialize $\theta^{(0)}$, sets $S_r \leftarrow \emptyset \quad \forall r < R$
    \For{$t = 0, 1, \dots$}
      \State \textbf{Data Parallel For} $r \in \{0, \dots, R-1\}$:  
      \State \quad $\theta_r \leftarrow \theta_r - \eta \nabla \mathcal{L}_r(\theta_r)$; \Comment{Local Update}
      \State \quad $S_r \leftarrow S_r \cup \mathrm{supp}(\Delta \theta_r)$; \Comment{Track modified indices}
      \If{$(t+1) \bmod T_{\mathrm{sync}} = 0$}
        \State $\mathcal{I}_{\mathrm{all}} \leftarrow \bigcup_{r} S_r$; \Comment{Global changed indices}
        \State \textbf{Sync} $\theta$: $\forall i \in \mathcal{I}_{\mathrm{all}}$, update $\theta[i]$: \Comment{Priority Merge}
        \State \quad $\theta[i] \leftarrow \theta_{k}[i] \quad \text{where } k = \max \{r \mid i \in S_r\}$
        \State Broadcast $\theta$ to all ranks; reset $S_r \leftarrow \emptyset$;
      \EndIf
    \EndFor
  \end{algorithmic}
\end{algorithm}


\section{EVALUATION}
\label{sec:evaluation}
\subsection{Experiment Environment}\label{sec:eval:config}

\textbf{System Setup.}
Our system is implemented on TorchRec~\cite{torchrec}, a production-grade DLRM framework, ensuring our evaluation reflects industrial practices. We use Fbgemm~\cite{fbgemm} for high-performance embedding kernel operations and Gloo~\cite{gloo} for its efficient collective communication, which is critical for our custom synchronization protocol.

All experiments are conducted on an 8-node cluster designed to emulate a production inference environment. Each node is equipped with 4× NVIDIA H100 GPUs~\cite{nvidia_h100} (80 GB HBM3) to accelerate the compute-intensive dense layers of the DLRM. For our CPU-based co-located training, each node features a dual-socket AMD EPYC 9684X system~\cite{amd_epyc9684x} with a total of 12 TB DDR5 memory per node. Each CPU consists of 8 CCDs, each equipped with 96 MB of L3 cache (768 MB total). Although CCDs are not directly exposed as hardware NUMA nodes, we treat each CCD as a logical isolation unit to enable fine-grained NUMA-aware performance isolation.
Inference nodes are interconnected via a 100 Gbps InfiniBand EDR network, ensuring that performance bottlenecks are isolated to the node level rather than the network.

\textbf{Compared Baselines.}
We evaluate \sysname{} against a range of established strategies to comprehensively analyze the performance-accuracy trade-off across design spaces. To stress-test compared solutions, we set the P99 latency constraint to 10ms.
\begin{itemize}[leftmargin=*]
    \item \emph{NoUpdate}, which serves as the lower-bound baseline for accuracy and the upper-bound for performance. By never updating the model, it introduces zero update overhead but enormous accuracy degradation due to staleness.
    \item \emph{DeltaUpdate (Streaming Update)}~\cite{xie2020kraken,liu2022monolith,sima2022ekko}, which represents the industry-standard practice for balancing freshness and overhead, as depicted in \autoref{fig:dlrm_system}.
    It synchronizes only changed parameters (deltas) rather than full models, reducing network traffic. We implement this to reflect the best-case scenario of the conventional decoupled training-inference paradigm.
    \item \emph{QuickUpdate}~\cite{matam2024quickupdate} represents the state-of-the-art in communication-efficient updates within the decoupled paradigm. It optimizes the Delta Update approach by applying a gradient-magnitude filter, transmitting only the top 5\% or 10\% most significant parameters. 
    This baseline is crucial for demonstrating that prior optimizations, while reducing network traffic, are insufficient because they (a) still incur inter-cluster latency and (b) introduce accuracy loss from dropped updates. 
\end{itemize}


\begin{table}[t]
\centering
\caption{Datasets for accuracy \& performance testing}
\label{tab:datasets_all}
\resizebox{\linewidth}{!}{
\begin{tabular}{lcc}
\hline
\textbf{Dataset}         & \textbf{Dataset Size (Samples)} & \textbf{Embedding Table Size} \\
\hline
Avazu                   & 4.7GB (32.3M)                    & 0.55 GB                        \\
Criteo           & 11GB (45.8M)                     & 1.9 GB                         \\
BD-TB                    & 1.5TB (5B)                       & 50 TB                          \\
Avazu-TB                & 0.72TB (5B)                      & 50 TB                          \\
Criteo-TB        & 1.2TB (5B)                       & 50 TB                          \\
\hline
\end{tabular}
}
\end{table}

\textbf{Benchmark Datasets.}
Our evaluation employs a combination of public and production-scale datasets to rigorously assess both the algorithmic correctness and systems performance of \sysname{} under realistic conditions. Key characteristics of all datasets are summarized in Table~\ref{tab:datasets_all}.
\begin{itemize}[leftmargin=*, noitemsep, topsep=0pt, parsep=0pt, partopsep=0pt]
    \item \emph{Accuracy-Centric Evaluation}: We evaluate recommendation quality on two standard public benchmarks—Avazu~\cite{avazu} and Criteo~\cite{criteochallenge}—which offer well-established ground truth for accuracy assessment. And then further include a large-scale production dataset (BD-TB) to validate our results in a real-world environment.
\item \emph{Systems-Centric Evaluation}: To stress-test the scalability and efficiency of our system under production-scale constraints, we synthetically scale the public datasets to 50TB, matching the scale of our industrial BD-TB dataset. This scaling is critical as the primary systems bottlenecks (memory capacity, bandwidth, update overhead) only manifest at this massive scale. These scaled datasets are generated using modified NVIDIA DLRM synthesis scripts~\cite{nvidia_dlrm_scripts}, calibrated to emulate real-world traffic patterns: a sustained load of ~100 million (±5\%) requests every 5 minutes, generating ~25GB of new training data. This models the continuous, high-volume update stream of a major online platform.
\end{itemize}



\textbf{Evaluation Metrics.}  
We evaluate \sysname{} along three main dimensions to capture its overall efficacy: (1) \emph{Model Update Cost}, measured as the time overhead required to synchronize new model updates. This is the primary metric for our claim of ``near-zero-overhead'' updates. (2) \emph{Recommendation Accuracy}, measured by the Area Under the ROC Curve (AUROC) to evaluate model quality.
(3) \emph{End-to-End Performance}, assessed using P99 latency (99th percentile) to reflect tail behavior under load.
These metrics collectively quantify the trade-off between freshness, accuracy, and overhead.

\subsection{Model Update Cost Results}
\label{sec:eval-performance}

\begin{figure*}[hbt]
  \centering
  \begin{subfigure}[t]{0.3\linewidth}
    \centering
    \includegraphics[width=\linewidth]{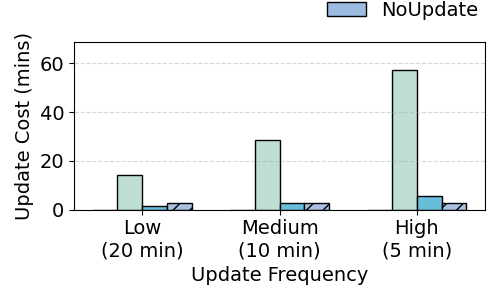}\vspace{-2ex}
    \caption{Avazu-TB}
    \label{fig:update-cost-avazu}
  \end{subfigure}
  \begin{subfigure}[t]{0.3\linewidth}
    \centering
    \includegraphics[width=\linewidth]{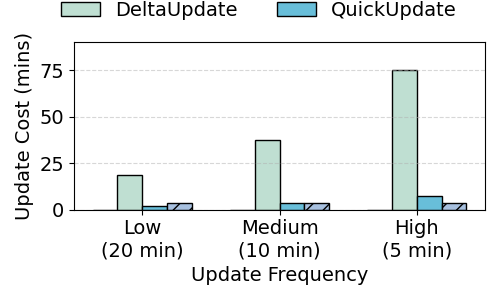}\vspace{-2ex}
    \caption{Criteo-TB}
    \label{fig:update-cost-criteo}
  \end{subfigure}
  \begin{subfigure}[t]{0.3\linewidth}
    \centering
    \includegraphics[width=\linewidth]{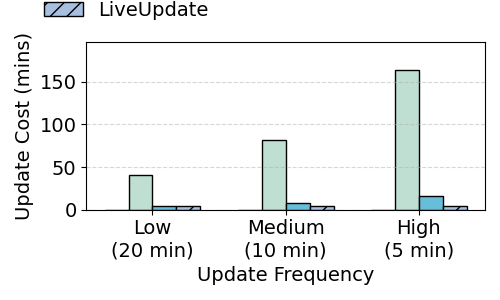}\vspace{-2ex}
    \caption{BD-TB}
    \label{fig:update-cost-X}
  \end{subfigure}
  \caption{Update costs of different methods across production-scale datasets}
  \label{fig:update-cost}
\end{figure*}

\autoref{fig:update-cost} compares the total update cost of different model update strategies across three datasets (Avazu-TB, Criteo-TB, and BD-TB) under varying 
update frequencies (20-minute, 10-minute, and 5-minute intervals, representing low, medium, and high update frequencies, respectively) in an hour. For example,
5-minute intervals represent updating the model every 5 minutes. And the update cost in \emph{DeltaUpdate} and \emph{QuickUpdate} represents the total model transmission time in an hour. The \sysname{} update cost represents the total training time in an hour because it does not require model transmission.

\emph{NoUpdate} introduces zero update cost since no model updates are performed. \emph{DeltaUpdate} incurs the highest overhead, with total synchronization time exceeding 60 minutes for a 5-minute update frequency on Avazu-TB. This confirms that full delta synchronization is prohibitively expensive at high frequencies, making it impractical for production systems. \emph{QuickUpdate} reduces this cost by transferring only a subset of parameters (5-10\%), making it more efficient at low frequencies (20-minute intervals). However, its cost scales linearly with update frequency, as the absolute volume of transmitted data remains significant.

In contrast, \sysname{}'s cost is dominated by local computation and is largely independent of update frequency. It consistently achieves the lowest overhead at high update frequencies, reducing update time by up to 2× compared to \emph{QuickUpdate}. For example, at the 5-minute interval, the update cost of \sysname{} is only 3-5 minutes. This result directly demonstrates the fundamental advantage of our architecture: by eliminating inter-cluster network transmission, the cost of freshness is decoupled from the update frequency.




\vspace*{-1ex}
\subsection{Recommendation Accuracy Comparison}
\label{sec:eval-accuracy}

We evaluate all methods with a 10-minute update window and report average AUC over a 1-hour period. Each training round starts from the same Day-1 checkpoint, with hourly full model updates using the previous hour's data, followed by evaluation in the next hour using a 10-minute sliding window. Our comparison includes \emph{QuickUpdate-$\alpha$\%} (top $\alpha$\% gradients), \sysname{}-$\alpha$ (LoRA with fixed rank $\alpha$), and \sysname{} (dynamic LoRA rank). Detailed AUC results are in \autoref{tab:abtest-accuracy-multi}, with the following observations:

\begin{table}[t]
\centering
\caption{Average AUC improvement (\%) over 1 hour with 
10-minute update intervals on three datasets.}
\label{tab:abtest-accuracy-multi}
\begin{tabular}{llll}
\hline
\textbf{Update Strategy} & \textbf{Avazu} & \textbf{Criteo} & \textbf{BD-TB}  \\ \hline
DeltaUpdate         & 0              & 0                      & 0              \\
NoUpdate                & -0.19          & -2.24                  & -1.29          \\
QuickUpdate-5\%          & -0.05          & -0.06                  & -0.07          \\
QuickUpdate-10\%         & -0.05          & -0.04                  & -0.03          \\
\sysname{}-8 (1\% extra Mem.)              & +0.07          & +0.09                  & +0.04          \\
\sysname{}-16/64 (4\% extra Mem.)          & +0.13          & +0.24                  & +0.14          \\
\textbf{\sysname{} (2\% extra Mem.)}    & \textbf{+0.09} & \textbf{+0.20}         & \textbf{+0.12} \\
 \hline
\end{tabular}
\end{table}
The \textit{DeltaUpdate} serves as the baseline (0\%), while \textit{NoUpdate} yields the lowest AUC (-0.19\% to -2.24\%), highlighting the need for regular model updates. \textit{QuickUpdate} mitigates degradation as update size increases (5\% to 10\%), but still underperforms compared to the baseline, suggesting limited benefit from partial updates.
In contrast, all {\sysname{}} variants outperform the baseline (+0.04\% to +0.24\%). {\sysname{}-16/64} achieves the highest gains (+0.13\% to +0.24\%) with 4\% memory overhead. The standard {\sysname{}} offers better tradeoff (+0.09\% to +0.20\%) with just 2\% memory cost.

\begin{figure}[t]
  \centering
  \includegraphics[width=0.9\linewidth]{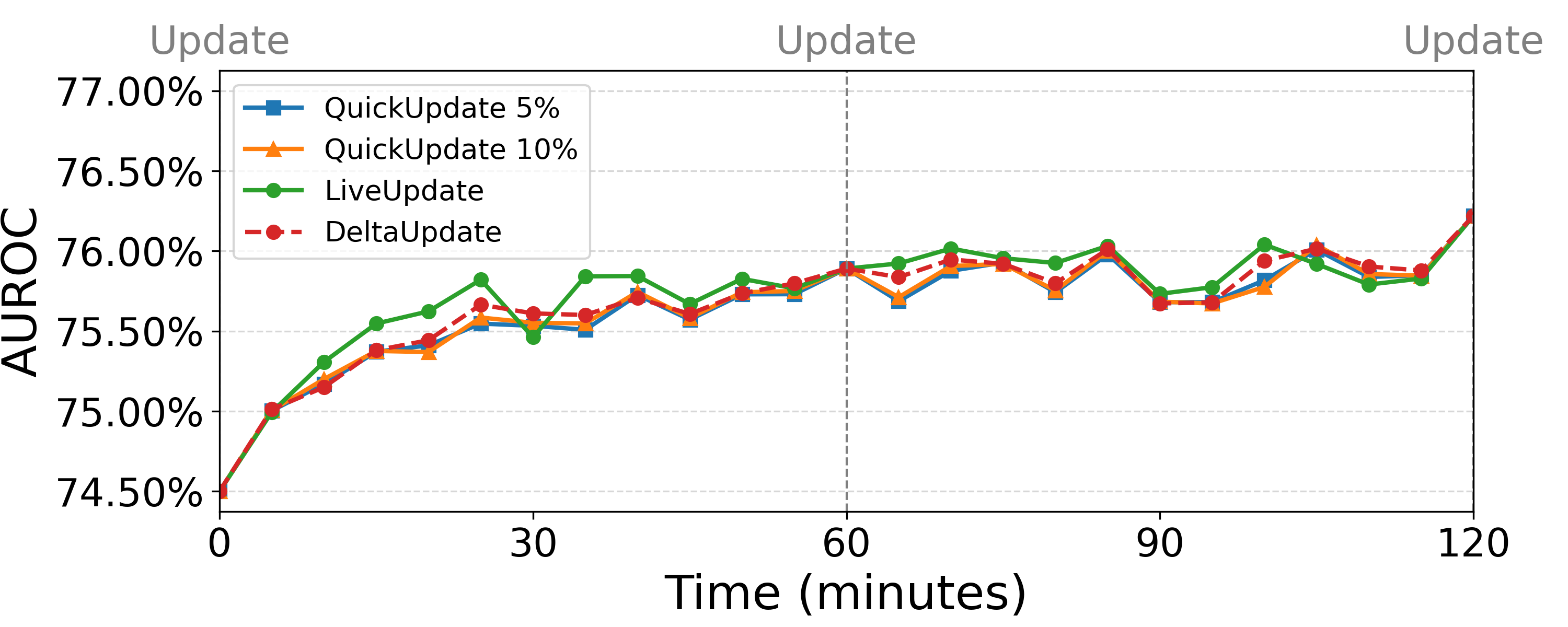}
  \caption{\xg{Accuracy changes in BD-TB dataset during two hours. Grey vertical lines indicate hourly full-parameter update.}}
  \label{fig:overall-update}
\end{figure}

To further analyze the accuracy changes over time, we plot the AUC changes in the BD-TB dataset over two hours in \autoref{fig:overall-update}. \xg{In this experiment, all methods update every 5 minutes. \textit{QuickUpdate} and \sysname{} further apply hourly full-update to mitigate error accumulation due to their reduced parameter update sizes. We ignore the update cost (e.g., data transfer time from training cluster to inference cluster) to focus solely on accuracy.
We have several interesting observations. First, although \sysname{} avoids frequent inter-cluster parameter transfer, it surpasses \textit{DeltaUpdate} (full updates) on accuracy most of the time, demonstrating the effectiveness of LoRA training. Second, the accuracy gap between \sysname{} and \textit{DeltaUpdate} narrows from 45min as errors accumulate, then widens after the full-update synchronization at 60min, showing the effectiveness of our tiered update strategy.
Lastly, due to the filtered parameters, the accuracy of \textit{QuickUpdate} is slightly below \textit{DeltaUpdate} (equivalent to \textit{QuickUpdate}-100\%), necessitating the design of this work. 
}
In addition, industry studies (ByteDance~\cite{shanprimus}, Tencent~\cite{lin2024understanding}) show even 0.03\%--0.07\% AUC gains can boost revenue by 0.4\%--2.4\%. Thus, \sysname{}’s improvements could translate to +1.60\%--4.11\% revenue gains (potentially tens of millions of dollars) based on industry conversion rates reported.

\subsection{End-to-End Performance Results}
\label{sec:performance}
\begin{figure}[t]
  \centering
  \includegraphics[width=0.9\linewidth]{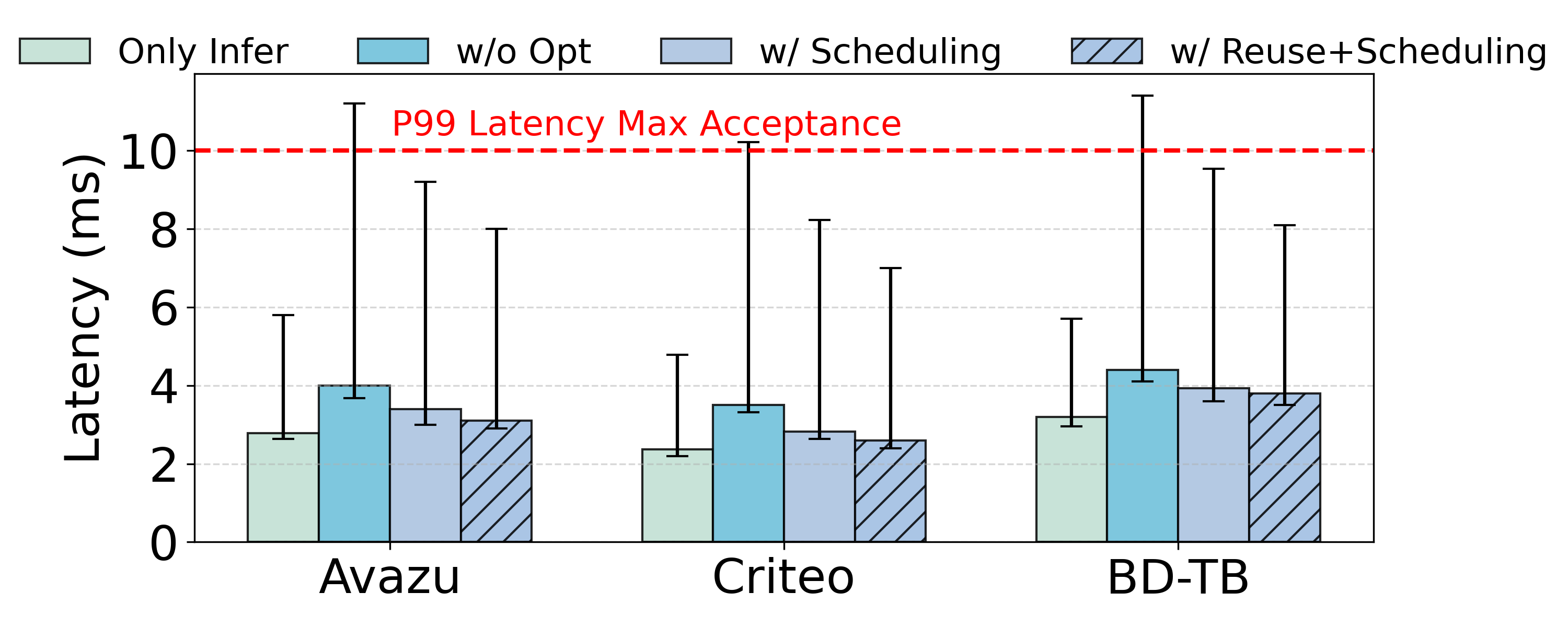}\vspace{-1ex}
  \caption{Effectiveness of performance isolation techniques}
  \label{fig:latency-ablation}
\end{figure}
We evaluate the impact of \sysname{} on the critical metric of inference tail latency (P99) under concurrent training load. \autoref{fig:latency-ablation} presents an ablation study comparing the latency of our full system against progressively optimized configurations. ``Only Infer" is the lower bound (no training), while ``w/o Opt" is naive co-location without optimizations (upper bound). 

The results demonstrate the incremental effectiveness of our optimizations: 1) 
No optimization (w/o Opt) causes a significant latency increase, confirming the severe interference problem.
2) NUMA-Aware Scheduling (w/ Scheduling) isolates training and inference workloads onto dedicated NUMA domains, drastically reducing cache and memory bandwidth contention. The resulted latency satisfies the P99 requirement.
3) Adding our embedding vector reuse mechanism (w/ Reuse+Scheduling) further reduces latency by eliminating redundant memory accesses, transforming the training thread's memory pattern from write-intensive to read-friendly.
Importantly, the full optimization results in performance nearly indistinguishable from the Inference Only lower bound, demonstrating that our co-location strategy achieves near-zero latency overhead.
\subsection{Ablation Study on Runtime Overhead}
\label{sec:V-E}
We conduct a detailed ablation study to quantify the runtime overhead of \sysname{}, focusing on memory consumption, energy usage, and CPU utilization.

\textbf{Memory Consumption.}
\begin{figure}[t]
  \centering
  \includegraphics[width=0.9\linewidth]{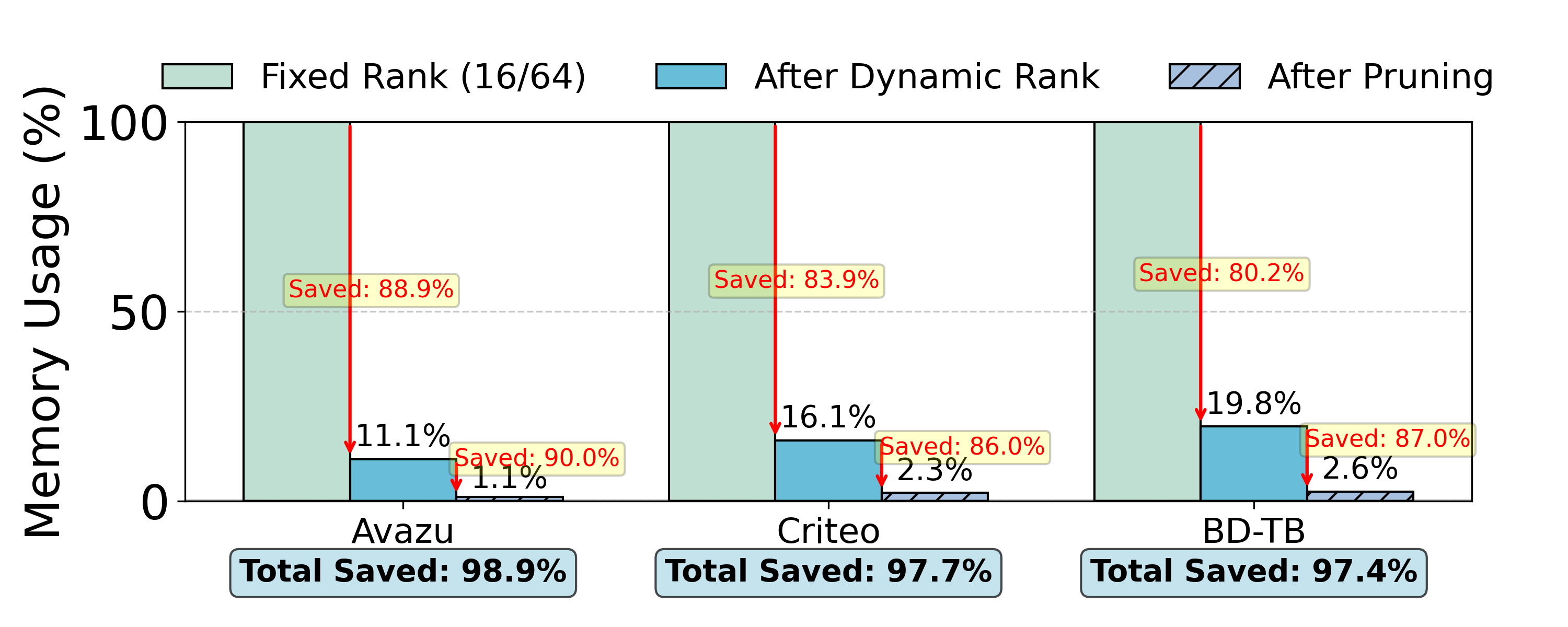}
  \caption{Effectiveness of memory optimization techniques}
  \label{fig:lora-memory-ablation}
\end{figure}
In memory management, we have developed two key optimizations: dynamic rank adaptation and parameter pruning. \autoref{fig:lora-memory-ablation} illustrates their dramatic impact on memory usage. The Fixed Rank approach serves as our baseline, representing memory usage with standard fixed LoRA ranks (e.g., 16 or 64). Dynamic rank adaptation alone achieves extraordinary savings between 80\% to 89\% across all datasets. This substantial improvement stems from the model's ability to intelligently adjust LoRA rank based on training data characteristics, eliminating unnecessary memory allocation. Building on this foundation, our pruning optimization cuts memory usage by an additional 90\%, achieving a total reduction of 97–99\%. This reduction stems from pruning parameters with negligible impact, enabling more efficient memory use. 
For a large-scale model of a 50 TB LoRA module, these savings correspond to 
reducing the memory footprint to roughly 1--3\% of the original size,
which equals about 0.5 to 1.5 TB. This substantial reduction makes 
training such massive models is more feasible and resource-efficient
without sacrificing accuracy.

\textbf{Energy and CPU Utilization.}
\begin{figure}[t]
  \centering
  \begin{subfigure}[t]{0.48\linewidth}
    \centering
    \includegraphics[width=\linewidth]{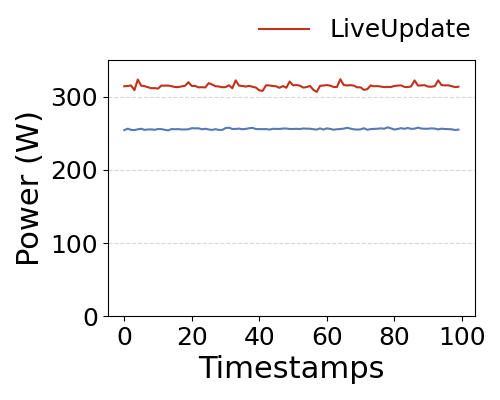}
    \caption{Power Consumption}
    \label{fig:energy}
  \end{subfigure}
  \begin{subfigure}[t]{0.48\linewidth}
    \centering
    \includegraphics[width=\linewidth]{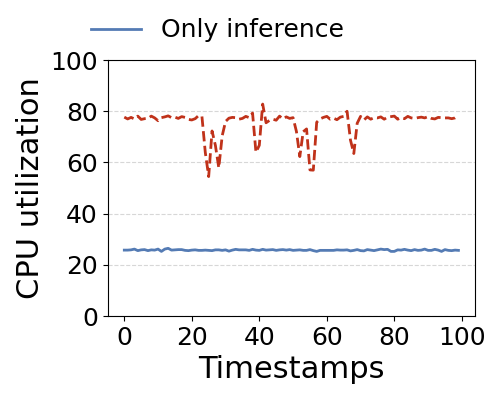}
    \caption{CPU Utilization}
    \label{fig:cpu-util}
  \end{subfigure}
  \caption{CPU power and utilization before and after applying \sysname{}}
  \label{fig:cpu-power-util}
\end{figure}
While introducing LoRA training increases energy consumption compared to inference-only operation, the absolute overhead remains modest, as shown in \autoref{fig:energy}. 
Critically, this minor energy cost is economically justified: while Intel reports that typical data center energy optimizations save ``tens of thousands of dollars'' annually~\cite{intel_dcm}, our system's 0.04–0.24\% revenue gain (from accuracy improvements) represents millions in potential upside, far outweighing the operational expense. Furthermore, as shown in \autoref{fig:cpu-util}, \sysname{} utilizes previously idle CPU cycles for training, increasing hardware efficiency without degrading inference performance (GPU inference time P99 latency$<$10 ms). This transforms wasted capacity into a valuable resource for model freshness.
\subsection{Scalability}
\label{sec:eval-scalability}
\begin{figure}[!t]
  \centering
  \includegraphics[width=0.8\linewidth]{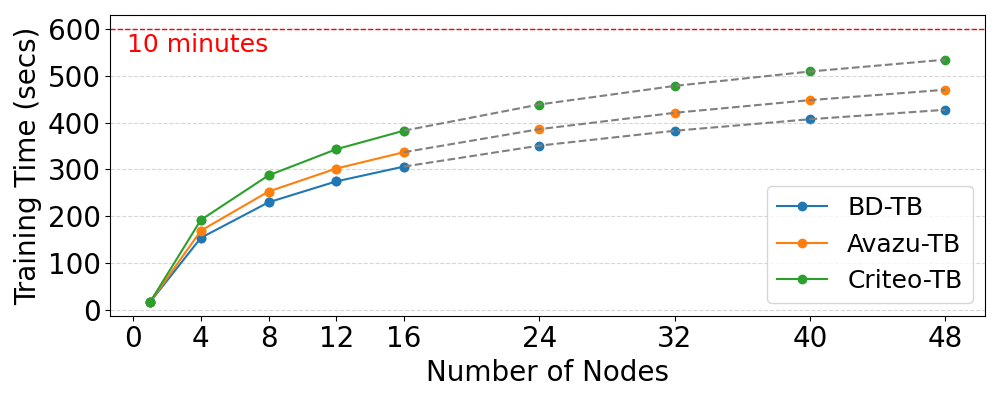}\vspace{-1ex}
  \caption{Scalability of \sysname{} with the increase of inference node numbers}
  \label{fig:train_time}
\end{figure}
We evaluate the scalability of \sysname{} by scaling the inference cluster to 16 nodes, representing an enterprise-grade deployment. \autoref{fig:train_time} shows the growth in training time as the number of nodes increases.

The results demonstrate that synchronization time scales logarithmically ($O(\log N)$) with the number of nodes. This favorable scaling is attributed to our use of Gloo's tree-based AllGather collective~\cite{gloo}, which avoids the linear scaling of naive synchronization schemes. While communication becomes the dominant cost at scale, the efficient algorithm ensures this overhead grows manageably.

Based on this observed logarithmic trend, we project performance for larger clusters of 24 to 48 nodes (simulated, dashed line in \autoref{fig:train_time}). The projection shows that the training time remains under 10 minutes, comfortably meeting the strict freshness requirements of production recommendation systems. This confirms that \sysname{}'s architecture is practical for large-scale, real-world deployment.
\xg{Note that industrial practice usually generates recommendations by combining outputs from multiple DLRM models (e.g., dislikes, trending, long-term preferences), with each model deployed on its own moderate-scale cluster. Thus, 48 nodes is a reasonable scale to reflect real-world scalability requirement.}

\section{RELATED WORK}
\label{sec:related-work}  
\par\textbf{Model Update Strategies in Real-Time DLRMs.}
\xg{Although prior works~\cite{zhao2023recd, tyagi2023accelerating, sethi2022recshard, miao2021het} have optimized training cluster bottlenecks (e.g., memory management, partitioning), they overlook the critical challenge of efficiently synchronizing updated multi-terabyte models to inference clusters. 

The prevailing solution for this online update problem involves delta-based strategies~\cite{xie2020kraken, liu2022monolith, sima2022ekko},} 
where modified parameters must be periodically transmitted from training nodes to inference nodes, resulting in significant network overhead. 
Ekko~\cite{sima2022ekko} further optimizes bandwidth consumption by exploiting spatial locality in parameter access patterns. HugeCTR~\cite{wei2022gpu} introduces the specialized HPS storage architecture to accelerate updates. These methods still fundamentally depend on full parameter updates, which can be costly in terms of both time and network resources.
QuickUpdate~\cite{matam2024quickupdate} performs partial updates by sending only the top 10\% most significant parameter changes, supplemented by occasional full updates. Although this reduces data volume, it still inherits the DLRM serving architecture’s heavy embedding-table transfers across the network. In contrast, we shift the paradigm by removing the update pipeline altogether: instead of sending parameters between training and inference systems, we place an adaptation module directly on inference nodes, enabling real-time updates with no network transfer overhead.

\par\textbf{Low-Rank Adaptation.}
Low-rank adaptation has been a key technique for reducing the computational and communication overhead in distributed machine learning systems. LoRA~\cite{hu2022lora} extends this idea to large language models (LLMs), enabling efficient fine-tuning with only a small number of trainable parameters.
FedPara~\cite{hyeon2021fedpara} introduces a Hadamard-based 
low-rank reparameterization to reduce communication costs in federated learning.
Nguyen et al.~\cite{nguyen2024towards} leverage LoRA in edge-based recommendation systems, reducing update traffic while preserving user privacy. 
In contrast, we apply LoRA to streaming DLRMs with novel modifications that enable truly real-time adaptation on inference nodes, maintaining accuracy without network transmission overhead.

\section{CONCLUSION}
\label{sec:conclusion}
This work addresses the core tension between model freshness and synchronization overhead in industrial DLRMs. Traditional decoupled architectures suffer from excessive delays when transferring EMT updates, leading to accuracy degradation and revenue loss. \sysname{} introduces a paradigm shift by co-locating lightweight LoRA trainers within inference clusters, leveraging underutilized CPU resources and the inherent low-rank structure of gradients. Our approach integrates two core innovations: dynamic rank adaptation via singular value pruning to maintain accurate updates with bounded memory overhead; and contention-free training management through NUMA-aware isolation, data-reuse method and load-sensitive scheduling to meet strict latency requirements. In production-scale evaluations, \sysname{} reduced synchronization costs by 2$\times$ and achieved sub-second model freshness—while improving recommendation accuracy by 0.04--0.24\% within freshness-critical windows and incurring negligible latency overhead. These results demonstrate that localized updates can effectively eliminate the staleness bottleneck without compromising serving performance.
\section*{Acknowledgements}

This work is supported by Guangdong and Hong Kong Universities ``1+1+1'' Joint Research Collaboration Scheme (project No. 2025A0505000012) and a research fund from ByteDance.


\bibliographystyle{IEEEtranS}
\bibliography{refs}

\end{document}